\documentclass[fleqn,usenatbib]{mnras}
\usepackage{newtxtext,newtxmath}
\usepackage[T1]{fontenc}
\DeclareRobustCommand{\VAN}[3]{#2}
\let\VANthebibliography\thebibliography
\def\thebibliography{\DeclareRobustCommand{\VAN}[3]{##3}\VANthebibliography}

\usepackage{graphicx} 
\usepackage{amsmath}
\usepackage{float}
\usepackage{url}
\usepackage{subcaption}
\usepackage{caption}
\usepackage{hyperref}
\usepackage{tikz}
\usepackage{fancyhdr}
\usepackage{cleveref}
\usepackage{xspace}
\usepackage{threeparttable}
\usepackage{longtable}
\usepackage{array}
\usepackage{makecell}
\usepackage{placeins}
\usepackage[dvipsnames]{xcolor}
\usepackage{ulem}

\newcommand{\orcidicon}[1]{%
  \href{https://orcid.org/#1}{\includegraphics[height=8pt]{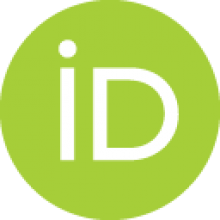}}%
}

\crefname{equation}{}{}  
\Crefname{equation}{}{}  

\title[\spine: Symbolic $P(k)$ Inference Emulator]{\spine: Symbolic Models to Predict the Evolution of the $\Lambda$CDM Nonlinear Power Spectrum}

\author[M. Chauhan et al.]{
Manvi Chauhan\,\orcidicon{0009-0000-6339-5415}$^{1,2}$\thanks{E-mail: m.chauhan-2018@hull.ac.uk}
Daniel J. Farrow\,\orcidicon{0000-0003-2575-0652}$^{1,2}$ 
Ariel G. S\'anchez\,\orcidicon{0000-0003-1198-831X}$^{3,4}$ 
Kevin Pimbblet\,\orcidicon{0000-0002-3963-3919}$^{1,2}$
Marika Asgari\,\orcidicon{0000-0002-3064-083X}$^{5}$  \and
David M. Benoit\,\orcidicon{0000-0002-7773-6863}$^{1}$
\\
$^{1}$E. A. Milne Centre for Astrophysics, University of Hull, Cottingham Road, Hull, HU6 7RX, UK
\\
$^{2}$Centre of Excellence for Data Science, AI, \& Modelling (DAIM), University of Hull, Cottingham Road, Hull, HU6 7RX, UK
\\
$^{3}$Max-Planck-Institut f\"{u}r extraterrestrische Physik, Postfach 1312, Giessenbachstr., 85748 Garching, Germany
\\
$^{4}$Universit\"{a}ts-Sternwarte M\"{u}nchen, Fakult\"{a}t f\"{u}r Physik, Ludwig- Maximilians-Universit\"{a}t M\"{u}nchen, Scheinerstrasse 1, 81679 M\"{u}nchen, Germany
\\
$^{5}$School of Mathematics, Statistics and Physics, Newcastle University, Herschel Building, NE1 7RU, Newcastle-upon-Tyne, UK
}

\pubyear{\the\year{}}

\newcommand{\LHmapeSPINE}{5.06}
\newcommand{\FIDmapeSPINE}{1.73}
\newcommand{\PLANCKmapeSPINE}{1.92}

\newcommand{\LHmapeSPINEX}{3.61}
\newcommand{\FIDmapeSPINEX}{1.35}
\newcommand{\PLANCKmapeSPINEX}{2.38}

\newcommand{\spine}{\texttt{SPINE}\xspace}
\newcommand{\spinex}{\texttt{SPINEX}\xspace}

\begin{document}

\label{firstpage}
\pagerange{\pageref{firstpage}--\pageref{lastpage}}
\maketitle

\begin{abstract}
We present \spine and \spinex, a pair of analytical emulators developed to predict the nonlinear power spectrum based on its linear counterpart and several essential cosmological parameters within the range of $0.01\;h\;\mathrm{Mpc}^{-1} <k< 2\;h\;\mathrm{Mpc}^{-1} $. The primary difference between the two models is their parameterisation. Our methodology is grounded in the mapping originally proposed by Peacock and Dodds (1996). Both models are defined by clear mathematical expressions derived from symbolic regression, a machine learning technique that utilises genetic programming to identify analytical equations that accurately represent the underlying data. This approach provides a more interpretable and efficient alternative to conventional numerical methods that should also exhibit superior extrapolation behaviour beyond the training range. The emulators have been trained on the $\Lambda$CDM Quijote Latin Hypercube simulations. We provide fits for both emulators across three distinct scenarios: cosmologies within 20$\sigma$ of the Planck-2018 observations, a designated fiducial cosmology, and a broader range of cosmologies sampled from the Quijote simulations. Our findings indicate that \spine and \spinex maintain an accuracy of better than 5\% in the majority of cases. These emulators provide a rapid alternative to numerical methods, and future initiatives will focus on developing expressions that incorporate galaxy bias and redshift-space distortions. This advancement aims to enhance the modelling of redshift space power spectra across multiple redshifts, enabling their application in large-scale cosmological surveys.
\end{abstract}


\section{Introduction}
The large-scale structure of the universe observed today originated from primordial density fluctuations generated during the inflationary epoch. These density fluctuations, which were highly Gaussian in nature, grew through the gravitational instability of dark matter fluctuations. These evolved under cosmic expansion, forming the cosmic web of galaxies. The clustering of matter produces a significant statistical signature that encodes a wealth of information about the most fundamental aspects of our Universe. For instance, the clustering amplitude as a function of scale depends on the amounts of dark matter, baryons, and photons in our cosmological model, as well as the conditions of the primordial universe \citep{peebles_1980, klypin_lss, springel_lss, liddle_inflation, moderncosmo, desi_2024, kids_legacy}. In Fourier space, this statistic is represented by the power spectrum, $P(k)$, which serves as a fundamental statistic for constraining cosmological models.

On large scales, different Fourier modes evolve independently and can be efficiently described analytically using linear perturbation theory to produce a linear power spectrum of fluctuations \citep{peebles_1980}. However, perturbation theory is only valid up to scales of $k\lesssim0.2\;h\;\mathrm{Mpc}^{-1}$ and deviations can reach $20\%$ for scales below this \citep{RPT}. Models that predict linear spectra as functions of redshift and cosmological parameters are well-established. Some notable models include \cite{eisenstein1998,eisenstein1999}, which are accurate up to a few per cent. This category also includes Boltzmann equation solvers such as CAMB \citep{camb} and CLASS \citep{class}, which can produce the linear spectra of fluctuations for $\Lambda$CDM and alternate cosmologies with much accuracy.

The situation on small scales, however, becomes extremely complex. Highly nonlinear processes and gravity cause perturbations on various scales to couple. Gravitational collapse on these scales amplifies inhomogeneity, and perturbation theory can no longer predict the evolution of structure formation, thereby failing to capture the nonlinear power spectrum accurately.

Numerical N-body simulations are one of the methods that closely model the power spectrum deep in the nonlinear regime. However, these simulations are often time-consuming and resource-intensive. In the current age of cosmological experiments, large-scale structure surveys, such as \textit{Euclid} \citep{2024euclid, 2024euclid_2}, DESI \citep{desi_2024,desi_2025}, HETDEX \citep{hetdex}, and LSST/Rubin \citep{lsst_rubin} can constrain cosmological parameters with great accuracy. For this purpose, one requires fast calculation of the power spectrum and other clustering statistics. The growing interest in achieving swift calculations of the power spectrum, while considering alternatives to N-body simulations, has encouraged the exploration of other avenues that may effectively predict the nonlinear power spectrum. 

Surrogate methodologies are founded on the halo model \citep[see][for a review]{halomodel_review, halomodel_asgari}, which posits that the matter content of the universe is concentrated within dark matter halos. The Halofit method \citep{peacock_stableclustering, halofit_takhashi} proposes that the nonlinear power spectrum can be decomposed into a one-halo term—determined by the distribution of dark matter within individual halos—and a two-halo term, which describes correlations between particles residing in different halos. The free parameters of this model have been fitted to N-body simulations. HMCode \citep{mead_hmcode, hmcode}, a model derived from the halo model, computes the power spectrum by integrating various cosmological quantities, such as the halo mass function and the halo density profile. HMCode is also capable of modelling baryonic feedback effects. 

There are several advantages associated with the use of the halo model method over numerical emulation techniques. The halo model depends on symbolic expressions for its analysis of the power spectrum, making it more advantageous than emulation techniques in terms of its interpretability and extrapolation behaviour. Furthermore, expressions yield output that is less noisy when compared to numerical emulation methods, contributing to greater analytical clarity. However, it is also essential to recognise that even symbolic approaches necessitate the resolution of multiple complex integrals. In this case, running simulations remains a prerequisite as the free parameters are calibrated on N-body simulations. This may complicate their integration into current and future analysis pipelines. Despite various advantages, they do not achieve the same level of accuracy as numerical emulation techniques.

Numerous studies have begun to adopt machine learning (ML) techniques for their data analyses. ML tools have demonstrated great potential in addressing the challenges associated with current statistical methods. For instance, neural networks were used to constrain the background dynamics of the universe for various cosmological models \citep{cosmoNN1}; they were also utilised in \cite{cosmoNN2} to connect dark matter haloes to their density profiles. ML methods were also used to calibrate N-body simulations \citep{flamingo_NN}, in CMB experiments \citep{CMB_1} and even used for cosmological parameter inference \citep{NN_params, hubble_ml, NN_params2, nn_hubble}. ML emulators for the calculation of power spectra have also been developed, such as the EuclidEmulator \citep{euclid_emulator, euclid_emulator2} and the BACCO Emulator \citep{bacco1, bacco2}. Emulation techniques to predict the power spectrum often rely on Gaussian processes or neural networks and are trained on N-body simulations to predict the nonlinear power spectrum as a function of cosmological parameters \citep{heitmann_pk_emulator1,heitmann_pk_emulator2, euclid_emulator2, bacco1, bacco2, comet_emulator, cosmopower, zennaro2023bacco, aletheia}. 

Due to their intrinsic architecture, ML methods are frequently characterised as `black boxes.' This terminology arises from the complexity of these models, which renders their internal mechanisms and the reasoning behind the decisions made by the algorithms unexplainable to users. It is essential to utilise a model that has justifiable conclusions and possesses an architecture that is interpretable. By creating models that are explainable and interpretable, one can not only find correlations in the data, but by virtue of the transparent nature of such models, one can define the causal relationships derived from the insights learned from the data.

In the pursuit of obtaining simple, explainable methods that maintain speed and accuracy, many studies worked towards building a concise equation for the power spectrum. One of the first fitting formulae was described in a series of papers \citep{hklm, PD94, jain_mo_white, PD96}. This method relied on a specific mapping between the linear and nonlinear regimes—see Section \ref{sec:methodology_power_spectrum} for more details—alternatively, more recent studies have aimed at constructing analytical emulators for the power spectrum. These studies have effectively employed symbolic regression (SR) via genetic algorithms to derive straightforward mathematical functions that accurately describe various quantities of interest. This method evolves mathematical expressions to fit specific datasets and has experienced a surge in popularity in recent years, particularly within the realm of physics. Some examples include the rediscovery of physical laws \citep{miles_cranmer_orbital}, the investigation of galaxy size and formation \citep{sr_galaxy_size}, the classification of astronomical objects \citep{sr_object_classification}, and the analysis of the primordial power spectrum \citep{sr_primordial_spectra}. A few notable emulators for the power spectrum include \cite{syren-halofit, linear_bartlett, syren_new}, which approximate the linear and nonlinear power spectrum for $\Lambda$CDM and modified gravity models, respectively. \cite{orjuela2023,orjuela2024} produce approximations for the transfer function and have similar accuracy to the one described in \cite{eisenstein1999}. 
 
In this study, we present two emulators, \spine (Symbolic Power-spectrum INference Emulator) and \spinex (Symbolic Power-spectrum INference Emulator - X), which aim to produce the $\Lambda$CDM nonlinear power spectrum using straightforward mathematical expressions. The primary distinction between \spine and \spinex is the parameters that each model contains. We demonstrate that by integrating the mapping approach recommended in \cite{PD96}, it is feasible to derive a concise expression for the nonlinear power spectrum based on its linear counterpart and several critical cosmological parameters. We construct both emulators utilising SR, as implemented through PySR\footnote{\url{https://github.com/MilesCranmer/PySR.git}} \citep{pysr}.

The structure of this paper is as follows. In Section \ref{sec:data}, we delineate the data utilised for training and testing our model. Section \ref{sec:methodology} offers a theoretical foundation by providing a concise review of the power spectrum and the mapping outlined in \citep[see Section \ref{sec:methodology_power_spectrum}]{PD96}. This section further explores our analysis of the baryon acoustic oscillations (Section \ref{sec:methodology_bao}) and describes the SR methodology employed to derive our analytical expression (Section \ref{sec:methodology_sr}). We then describe the procedure by which one can obtain a prediction for the power spectrum using our equations (Section \ref{sec:methodology_calculate_pk}). In Section \ref{sec:results}, we provide details on the selection process of equations (Section \ref{sec:results_select_eqn}). This is succeeded by the symbolic expressions that make up \spine (Section \ref{sec:results_spine_eqn}) and \spinex (Section \ref{sec:results_spine_eqn}) and predictions for the nonlinear power spectrum across three distinct scenarios: cosmologies situated within 20$\sigma$ of the \cite{planck_2018} results (Section \ref{sec:results_planck}), a designated fiducial cosmology (Section \ref{sec:results_fiducial}), and cosmologies from the Quijote Latin hypercube (Section \ref{sec:results_LH}). We then discuss the other equations spawned by PySR (Section \ref{sec:results_auxilliary_eqns}). In Section \ref{sec:discussion}, we discuss the results in detail, providing a comparison between \spine and \spinex (Section \ref{sec:discussion_understanding}). Additionally, we evaluate the performance of the models (Section \ref{sec:discussion_model_performance}). We also conclude and outline future work in Section \ref{sec:conclusion}. 

\section{Data} \label{sec:data}
We employ the Quijote suite of N-body simulations \citep{quijote_simulations} to facilitate the training and evaluation of our model. This comprehensive suite comprises 44,100 N-body simulations, encompassing more than 7,000 distinct cosmological frameworks run using GADGET-III, an optimised version of GADGET-II \citep{gadget2}. All simulations are conducted within a cosmological volume of $1\; (h^{-1}\; \text{Gpc})^3$. Two primary sets of simulations are utilised in this research: the Latin hypercube (LH) suite and the Quijote fiducial cosmology set. LH sampling \citep{latin_hypercube,latin_hypercube2} is a critical method for tackling computationally intensive problems, as it facilitates the efficient exploration of extensive, multidimensional input spaces. This approach guarantees thorough coverage of each variable's range while utilising a reduced number of samples \citep{lh_sampling}.

The Quijote suite consists of 11,000 LH simulations. This dataset has been specifically developed for the purpose of training ML models. In our case, we used the fixed simulations set of 2000 $\Lambda$CDM realisations with a resolution of $(512)^3$ CDM particles. These simulations have a fixed initial random seed. The simulations are described by the five cosmological parameters -- $\left \{\Omega_\mathrm{m}, \Omega_\mathrm{b}, h, n_\mathrm{s}, \sigma_8\right\}$ with $\left\{M_{\nu}=0, w=-1 \right \}$ which denote the matter and baryon densities, the reduced Hubble constant ($h=H_0/100 \; \text{km} \;\text{s}^{-1} \; \text{Mpc}^{-1}$), the spectral index indicating the tilt of the primordial power spectrum, the root-mean-square density fluctuation when the linearly evolved field is smoothed with a top-hat filter of radius $8\; h^{-1}\mathrm{Mpc}$, the mass of neutrinos, and the equation of state parameter for dark energy, respectively. The limits of these parameters can be found in Table 1 of \cite{quijote_simulations}. The boundaries of the downsampled parameter space for both the training and test sets are given in Table \ref{tab:Quijote_limits}. 

To test our emulator, we also provide predictions for the power spectrum corresponding to cosmologies within 20$\sigma$ of the \cite{planck_2018} best-fitting $\Lambda$CDM model (hereafter Planck-2018). These cosmologies were selected from the LH. Moreover, any cosmology included in the Planck-2018 set was excluded from the test set to ensure the integrity of the analysis. We note here that this exclusion was not applied to the training set. Both \spine and \spinex were trained between the interval $k = 0.01 - 2\;h\;\mathrm{Mpc}^{-1}$ although \cite{quijote_simulations} state that at $z=0$, the simulations converged up to $k = 1\; h \;\mathrm{Mpc}^{-1}$ at 2.5\% for a fiducial resolution of $(512)^3$ CDM particles. For this reason, we calculate the performance metrics (see Section \ref{sec:methodology_sr}) up to this $k-$value. 

For our training and test set, we selected 200 random cosmologies from the extensive LH set. This selection of 200 was made to address constraints related to computational speed and memory resources. Furthermore, it was determined that this volume of data is adequate for our chosen ML algorithm to effectively identify an appropriate fitting function. We chose 65\% of the 200 random cosmologies, i.e. 130 cosmologies, as the training set, and the remaining 35\%, i.e. 70 cosmologies, make up the test set. However, after the removal of Planck-like cosmologies from the test set, 68 test cosmologies remained. Note here that the training sets for \spine and \spinex contain different sets of cosmologies (See Table \ref{tab:Quijote_limits}). The test set has been kept constant across both models to facilitate a reliable basis for comparative analysis. The test set comprises a well-balanced array of cosmologies that are both within and outside the training range of the emulators. This design enables us to evaluate the extrapolation behaviour of the models effectively.

Secondly, the Quijote fiducial simulations were used to test our emulator predictions for the fiducial cosmology. These simulations consist of paired fixed simulations (each independent pair using a different initial random seed, while the two members of each pair have fixed amplitudes and phases shifted by $\pi$), with parameter limits detailed in Table \ref{tab:Quijote_limits}. The values used are in good agreement with the results presented by Planck-2018. This simulation set encompasses 500 power spectra under 2LPT initial conditions and was conducted at the same resolution as the fixed LH cosmologies.

What makes fixed and paired-fixed simulations different from typical methods is the way initial conditions are generated \citep[see][for a review]{stats_paired_fixed}. Instead of sampling a Fourier space mode's amplitude from a Gaussian distribution, the amplitude is set to the root mean square value of the target power spectrum, thereby eliminating the intrinsic scatter at linear order. Fixed and paired simulations help to significantly reduce the impact of cosmic variance when modelling the evolution of density fluctuations \citep{pontzen_paired_fixed, pairedfixed, suppressing_cosmic_variance}.

\begin{table*}
\caption{Limits of the Quijote cosmologies that we use in this work. We utilised the Quijote LH cosmologies to train and test our model. The limits of the training and test sets for \spine and \spinex are given in columns marked as `Train' and `Test'. To obtain a prediction for Planck-like cosmologies, we chose 45 cosmologies from the LH that were within 20$\sigma$ of Planck-2018. Any Planck-like cosmologies from the test set have also been removed. Shown in the last column are the values for the Quijote fiducial cosmologies.}
    \centering
    \begin{tabular}{cccccccccc}
    \hline
    \textbf{Parameter} & \multicolumn{2}{c}{\textbf{Train (\spine)}} & \multicolumn{2}{c}{\textbf{Train (\spinex)}} & \multicolumn{2}{c}{\textbf{Test}} & \multicolumn{2}{c}{\textbf{Planck-like}} & \textbf{Fiducial} \\
                  & \textbf{Min} & \textbf{Max} & \textbf{Min} & \textbf{Max} & \textbf{Min} & \textbf{Max} & \textbf{Min} & \textbf{Max} & \\
    \hline
        $\Omega_m$ & 0.1019  & 0.4943  & 0.10170 & 0.4931  & 0.1087  & 0.495   &  0.1821 &  0.4503 & 0.3175 \\
        $\Omega_b$ & 0.03009 & 0.06995 & 0.03037 & 0.06995 & 0.03007    & 0.06955 & 0.04311 & 0.04311 & 0.049 \\
        $h$        & 0.5005  & 0.8877  & 0.5005  & 0.89870 &  0.5007    & 0.8987   & 0.57550 & 0.77310 & 0.6711 \\
        $n_s$      & 0.8005  & 1.1989  & 0.8005  & 1.1957  &  0.8013 & 1.1921  & 0.8867  & 1.04150 & 0.9624 \\
        $\sigma_8$ & 0.6001  & 0.9933  & 0.6009  & 0.9983  & 0.60307 & 0.9983  & 0.6965  & 0.9277  & 0.834 \\
        \hline
    \end{tabular}
    \label{tab:Quijote_limits}
\end{table*}

\section{Methodology} \label{sec:methodology}
In this section, we review the matter power spectrum and outline the methodology employed to derive its broadband shape. This includes the procedures implemented in the analysis of the baryon acoustic oscillation (BAO) features. Furthermore, we will introduce the ML algorithm utilised to generate a symbolic expression for the nonlinear power spectrum. 
\subsection{The Matter Power Spectrum} \label{sec:methodology_power_spectrum}
Consider a statistically homogeneous and isotropic density field $\rho(\textit{\textbf{r}})$. Overdensities in this field are described by $\delta(\textit{\textbf{r}}) \equiv \left(\rho(\textit{\textbf{r}}) - \overline{\rho}\right)/\overline{\rho}$ where $\overline{\rho}$ represents its mean value. The power spectrum, $P(k)$, of density fluctuations of such a field is then described as
\begin{equation}
    \langle \delta_{\mathrm{\mathrm{k}}} \delta_{\mathrm{k'}} \rangle = (2\pi)^3\delta_{\text{D}}(k+k')P(k)
\end{equation}

\noindent where $\delta_\mathrm{k}$ is the Fourier transform of the density contrast and $\delta_{\text{D}}$ is the Dirac delta function. Statistical isotropy requires $P(k)$ to be a function of only the magnitude of the wavevector $\textit{\textbf{k}}$, that is, $|\textit{\textbf{k}}| \equiv k$. The power spectrum is also the Fourier transform of the 2-point correlation function $\xi(r)$, which describes the excess probability of finding a pair of objects within a certain distance \textit{\textbf{r}} of each other with respect to a homogeneous distribution. It is given as
\begin{equation}
    \xi(r) = \langle \delta(\textit{\textbf{x}})\delta(\textit{\textbf{x}}+\textit{\textbf{r}})\rangle.
\end{equation}

\noindent $P(k)$ can then calculated as

\begin{equation}
    P(k) = \int \xi(r)e^{i \textit{\textbf{k}} \cdot \textit{\textbf{r}}} d^3r,
\end{equation}

\noindent such that $P(k)$ and $\xi(r)$ form a Fourier transform pair. The matter power spectrum represents a critical statistic for large-scale structure analysis, as it provides valuable insights into the clustering of matter at various scales and epochs. Additionally, it yields an intricate understanding of how cosmological parameters impact the process of structure formation.

On large scales, the power spectrum traces the spectrum of primordial density fluctuations and is therefore a direct probe into early universe physics. Given these initial conditions and cosmological parameters, the linear and quasi-linear power spectrum can be easily calculated through perturbation theory \citep[e.g.][]{perturb_theory, perturb_theory2}. However, on smaller scales, galaxy clustering and nonlinear gravitational effects affect the power spectrum in a non-trivial way, substantially enhancing the power on small scales compared to linear theory \citep{pk_large_to_small}. At these scales, it becomes essential to employ N-body simulations to model the late-time dynamics of the universe effectively. The outcomes derived from N-body simulations are then utilised to formulate phenomenological models and fitting equations of nonlinear gravitational clustering. 

\citet[hereafter PD96]{PD96} presented an analytical model that establishes a framework for mapping the statistics of both linear and nonlinear large-scale structures. This framework is built upon the scaling relations outlined in \cite{hklm} and can be understood conceptually via the spherical collapse model \citep[as described in detail in][]{PD94,peacock_stableclustering}. Their findings indicate that the volume-averaged nonlinear correlation function can be parametrised using the linear correlation function, contingent upon the nonlinear evolution resulting in a change of scale. This transformation of scales connects the linear scale $r_{\mathrm{L}}$ to the nonlinear scale $r_{\mathrm{NL}}$ through the established relation
\begin{equation}
    r_{\mathrm{L}} = \left [1+\bar{\xi}_{\mathrm{NL}}(r_{\mathrm{NL}}) \right]^{1/3} r_{\mathrm{NL}}.
    \label{eqn:hklm_relation}
\end{equation}

\noindent Where $\bar{\xi}_{\mathrm{NL}}(r_{\mathrm{NL}})$ is the nonlinear correlation function calculated at a nonlinear scale $r_\mathrm{NL}$. After this rescaling, the nonlinear correlation function would be a universal function of the linear one, 
\begin{equation}
    \bar{\xi}_{\mathrm{NL}}(r_{\mathrm{NL}}) = f \left[\bar{\xi}_{L}(r_{\mathrm{L}}) \right].
    \label{eqn:hklm_corr}
\end{equation}

\noindent PD96 proposed that a similar relationship may be established for the power spectrum, specifically in relation to its dimensionless version, $\Delta^2(k)$, which represents the fractional density variance per unit in ln$k$ or, simply stated, it tells us on what scales to expect significant density fluctuations. We calculate $\Delta^2(k)$ as
\begin{equation}
    \Delta^2(k) = \frac{k^3}{2\pi^2}P(k).
    \label{eqn:pk_conversion}
\end{equation}
\noindent The recipe of PD96 corresponds to the Fourier-space version of that of \cite{hklm}, resulting in the relation
\begin{equation}
    k_{\mathrm{L}} = \left [1+\Delta_{\mathrm{NL}}^2(k_{\mathrm{NL}}) \right]^{-1/3}k_{\mathrm{NL}},
    \label{eqn:pd_mapping_2}
\end{equation}
\noindent the phase space version of equation~\eqref{eqn:hklm_relation}. This then leads to the realisation that the nonlinear spectrum can be formulated as a universal function of the linear power spectrum
\begin{equation}
    \Delta_{\mathrm{NL}}^2=f_{\mathrm{NL}}\left[\Delta_{\mathrm{L}}^2(k_{\mathrm{L}})\right].
    \label{eqn:pd_mapping}
\end{equation}
\noindent PD96 calibrated the function $f_{\mathrm{NL}}$ by running a series of $\Lambda$CDM N-body simulations to isolate the functional form of $f_{\mathrm{NL}}$. The fitting formula proposed for the nonlinear power spectrum is based on a specific assumption: it posits that the nonlinear power spectrum is not influenced by the time evolution, i.e., the history of the linear power spectrum. Rather, it is determined exclusively by the linear power spectrum at the corresponding epoch. While their methodology provides a valuable framework for explanation, it lacks the necessary accuracy to effectively interpret current data.

Our work aims to exploit the mapping recommended by PD96 (equations~\ref{eqn:pd_mapping_2} and \ref{eqn:pd_mapping}) and predict the function $f_{\mathrm{NL}}$ using ML methods, specifically SR, to predict the redshift zero nonlinear power spectrum $P_{\mathrm{NL}}(k_{\mathrm{NL}})$ as a function of the linear power spectrum $P_{\mathrm{L}}(k_{\mathrm{L}})$ and cosmological parameters. We present two models

\begin{table}
\caption{The model parameters of each emulator. The left column contains the parameters that make up \spine. The right column contains the parameters that make up \spinex.}
    \centering
    \begin{tabular}{p{0.3\linewidth} p{0.3\linewidth}}
    \hline
    \textbf{\spine} - $\theta$ & \textbf{\spinex} - $\theta_X$\\
    \hline
        $h$ & $\omega_\mathrm{m}$ \\
        $\Omega_\mathrm{m}$ & $f_b=\omega_\mathrm{b}/\omega_\mathrm{m}$ \\
        $\Omega_\mathrm{b}$ & $n_\mathrm{s}$ \\
        $n_\mathrm{s}$ & $\sigma_{12}$ \\
        $n_\mathrm{L}=\left.\frac{d\ln P}{d\ln k}\right|_{k_\mathrm{L}/2}$ & $n_\mathrm{L}$ \\
        $\sigma_8$ &  $\tilde{x}$ \\
        $g_\mathrm{a}=D(a)/a$ &  \\
    \hline
    \end{tabular}
\label{tab:model_params}
\end{table}

\begin{enumerate}
    \item \spine, which predicts the $P_{\mathrm{NL}}(k_{\mathrm{KL}})$ in terms of $P_{\mathrm{L}}(k_{\mathrm{L}})$ and cosmological parameters $\theta$, where $\theta$ contains the parameters listed in Table \ref{tab:model_params}. All parameters are calculated at $z=0$. Here, $n_\mathrm{L}$ is the slope of the late-time linear power spectrum, $\sigma_8$ is the root-mean-square (RMS) of density fluctuations smoothed over spheres of radius $8 \; h^{-1}\;\mathrm{Mpc}$, and
    $g_\mathrm{a}=D(a)/a$ is the growth suppression factor where $D(a)$ is the linear growth factor, and $a$ is the scale factor.
    
    \item \spinex, which predicts the $P_{\mathrm{NL}}(k_{\mathrm{NL}})$ in terms of $P_{\mathrm{L}}(k_{\mathrm{L}})$ and cosmological parameters $\theta_X$, where $\theta_X$ contains the parameters listed in Table \ref{tab:model_params}. Here, $\omega_\mathrm{m}$ is the physical matter density, $f_\mathrm{b}$ is the baryon fraction, and $\sigma_{12}$ is the RMS of density fluctuations smoothed over spheres of radius 12 Mpc. $\tilde{x}$ \citep[see][]{aletheia}, is defined as 
    \begin{equation}
        \tilde{x}(\tau) = \int_{-\infty} ^{\tau} x(\tau')K(\tau - \tau'|\eta) d\tau',
        \label{eqn:xtilde}
    \end{equation}
    \noindent where 
    \begin{equation}
        x(z) = \frac{\Omega_\mathrm{m}(z)}{f^2(z)}.
    \end{equation}
    \noindent $\tilde{x}$ is an average of $x$ over past history using $\tau=\mathrm{ln}(\sigma_{12})$ as the time variable. $\tilde{x}$ encodes the cosmological dependence of the nonlinear evolution of the density field and parametrises the differences in the nonlinear power spectrum due to different growth histories. $K(\tau|\eta)$ is a Gaussian kernel with width $\eta$ which indicates memory of the nonlinear evolution. \cite{aletheia} found that $\eta=0.12$ provides an excellent characterisation of the deviations observed across the cosmologies and scales analysed. Therefore, we have chosen to adopt this value for $\eta$ in our work as well.

    In models of dynamical dark energy, cosmologies that possess identical linear power spectra can produce markedly distinct nonlinear structures \citep{dde_evolution}. The influence of dynamical dark energy on structure formation is contained within the parameter $\tilde{x}$. Consequently, it is expected that the \spinex is capable of producing reliable predictions for cosmologies that incorporate dynamic dark energy. However, the development and evaluation of a model for such cosmologies will be addressed in future work.

    The key difference between \spine and \spinex is in their input parameters. As noted by \cite{arguments_h, evolution_mapping}, the matter and baryon densities $\Omega_\mathrm{m}$ and $\Omega_\mathrm{b}$ have explicit $h$ dependencies. By using physical density parameters $\omega_\mathrm{m}$, the \spinex parameterisation tries to avoid this. The use of $h$-independent parameters enables a clearer examination of how $P(k)$ responds to variations in $h$ \citep{arguments_h}.  
    
\end{enumerate}

\noindent Concerning the function $f_{\mathrm{NL}}$, we chose to find a function of the form 
\begin{equation}
    \Delta_{\mathrm{NL}}^2(k_{\mathrm{NL}}) = \Delta_{\mathrm{L}}^2(k_{\mathrm{L}}) + N\left[\Delta_{\mathrm{L}}^2(k_{\mathrm{L}}),\left(\theta \;\text{or}\; \theta_X\right) \right],
    \label{eqn:func_form_SR}
\end{equation}

\noindent where $N$ is a function of the linear power spectrum and cosmological parameters of the model chosen. Equation~\eqref{eqn:func_form_SR} can be interpreted as representing the linear power spectrum, with $N\left[\Delta_{\mathrm{L}}^2(k_{\mathrm{L}}),\left(\theta \;\text{or}\; \theta_X\right) \right]$ functioning as an additional nonlinear correction similar to the one-halo term in the halo model. This interpretation facilitates an analysis of how nonlinearities impact the late-time power spectrum as a function of differing cosmological parameters. Equation~\eqref{eqn:func_form_SR} employs versions of the power spectrum from which the BAO signal has been removed. The treatment of the nonlinear evolution of the BAO occurs separately, and a detailed explanation of this process will be provided in the upcoming section.

\subsection{Baryon Acoustic Oscillations in the Power Spectrum} \label{sec:methodology_bao}
BAOs are oscillatory features in the power spectrum. These features had their origins before recombination when the ionised matter content of the universe was coupled to radiation through electromagnetic interaction. Gravitational attraction and the outward push of the photon pressure fought against each other, resulting in oscillations in the photon-baryon fluid. First identified in the galaxy density field by SDSS \citep{sdss_bao1} and 2dFGRS \citep{2dfgrs_cole}, BAOs function as a reliable standard ruler for quantifying the expansion of the universe. The distinctive oscillatory pattern of BAOs within the power spectrum facilitates the differentiation of the BAO signal from the general shape of the spectrum. This particular characteristic enables the effective isolation and removal of the BAO feature through a process known as `dewiggling'. Upon completion of this dewiggling process, the resulting power spectra are employed to fit an analytical model. This methodology enhances the accuracy with which SR can fit an equation for the power spectrum, as SR is only required to conform to the overall broadband shape of the dimensionless power spectrum.

As large-scale structures develop, matter particles move from their initial positions due to bulk flows and peculiar motions. These movements, which lead to the attenuation of the initially distinct BAO feature and the loss of some of its higher harmonic peaks, can be adequately described using a displacement field. This large-scale displacement represents the primary contributor to the decay of the matter propagator, a quantity that outlines the evolution of perturbations within the distribution of matter over time \citep{RPT, RPT2}. This concept is widely utilised in BAO reconstruction efforts within large-scale structure surveys, such as DESI \citep[see example][]{DESI_BAO_reconst} and even incorporated in models like HMCode \citep{hmcode}.

The magnitude of these displacements is quantified by the variance of the displacement field, $\sigma_{\mathrm{v}}^2$. This variance also serves as the one-dimensional velocity dispersion in the context of linear theory \citep{RPT}. Given that $\sigma_v^2$ measures the degree to which matter is displaced over time, an increased value of $\sigma_{\mathrm{v}}^2$ signifies a more pronounced damping effect on the BAO signal. We calculate this quantity as
\begin{equation}
    \sigma_{\mathrm{v}}^2 = \frac{1}{6\pi ^2} \int P(k)\;dk. 
\end{equation} 

\noindent This dampening effect presents itself as a broadening of the BAO peak in configuration space. In Fourier space, this represents a scale- and cosmology-dependent decrease in the amplitude of the oscillations in the power spectrum. These effects need to be modelled accurately in the linear power spectrum since the removal of the BAOs in its nonlinear counterpart relies on the linear one. The damping effect can be accurately modelled by splitting it as a sum of a smooth linear power spectrum and an oscillatory part that describes the impact of the BAO. This, therefore, takes the form of 
\begin{equation}
    P_{\mathrm{L}}^{\mathrm{dw}}(k) = P_{\mathrm{L}}^{\mathrm{nw}}(k) + \left [P_{\mathrm{L}}(k) - P_{\mathrm{L}}^{\mathrm{nw}}(k) \right]e^{-\frac{1}{2}k^2 \sigma_{\mathrm{v}}^2},
    \label{eqn:damped_linear_pk}
\end{equation}

\noindent where $P_{\mathrm{L}}(k)$ is the linear matter power spectrum and $P_{\mathrm{L}}^{\mathrm{nw}}(k)$ is the dewiggled linear power spectrum, which is obtained via a discrete sine transformation algorithm proposed in \cite{bao_dewiggling}. $P_{\mathrm{L}}^{\mathrm{dw}}(k)$ is now the linear power spectrum with damped BAO oscillations. The exponential component of equation~\eqref{eqn:damped_linear_pk} produces a Gaussian damping effect, which effectively models the attenuation of the BAO signal \citep{best_way_measure_bao,aletheia}. Given this prescription for dewiggling the linear power spectrum, we can obtain a smooth nonlinear power spectrum, $P_{\mathrm{NL}}^{\mathrm{nw}}(k)$, by doing\begin{equation}
    P_{\mathrm{NL}}^{\mathrm{nw}}(k) = P_{\mathrm{NL}}(k) \frac{P_{\mathrm{L}}^{\mathrm{nw}}(k)}{P_{\mathrm{L}}^{\mathrm{dw}}(k)},
    \label{eqn:nl_dewiggle}
\end{equation}

\noindent where $P_{\mathrm{NL}}(k)$ is the nonlinear power spectrum.  Additionally, the BAOs can be easily reintegrated into the power spectrum after the dewiggling has been completed, which can be achieved by rearranging equation~\eqref{eqn:nl_dewiggle} to obtain $P_{\mathrm{NL}}(k)$. Another motivating factor for dewiggling the power spectrum comes from \cite{PD94}, wherein the authors note that the mapping that forms the basis of their assumption fails for power spectra that have oscillations attributed to high baryon models, we therefore decided to remove this feature.  

\subsection{Symbolic Regression} \label{sec:methodology_sr}
SR involves the use of symbolic expressions for nonlinear regression, systematically integrating various mathematical operators to develop an optimal model \citep[see][]{GA_scientific_american, GA_haupt_haput, GA_overview_learning, GA_kronberger2024}. SR via genetic algorithms is a method inspired by the evolution of biological organisms. Following this, a group of individuals (represented by a family of equations) evolves to fit a dataset. Genetic algorithms incorporate two main elements: 
 
\begin{enumerate}
    \item A Darwinian concept of `fitness' operates to ensure that only the most effective equations advance, while less effective ones are systematically discarded. This process is governed by the fitness function, which evaluates both the loss and the complexity of the expressions involved.

    \item `Reproduction', where new individuals are spawned. This involves either the generation of new expressions via a combination of sub-expressions from the most successful individuals of the previous generation (crossover) or by randomly modifying an operator \citep[mutation, see][for details]{pysr}. These mutated individuals then replace the oldest members of the population, a process known as age-regularised evolution \citep{age_structure,age_fitness, age_regularized}.
\end{enumerate}

\noindent A notable advantage of using a compact mathematical expression to represent the power spectrum is the various benefits it provides when compared to numerical differential equation solvers, neural networks, or Gaussian emulators. As previously mentioned, symbolic expressions enhance the intuitive understanding of the underlying physics associated with power spectrum estimation. Furthermore, they are highly portable, as they can be implemented in the programming language of the user's preference and maintain their relevance even if the initial software becomes obsolete.

In this work, we make use of SR through PySR, an open-source library for SR in Python. PySR allows for massive parallelisation by dividing step (ii) into groups which evolve independently with occasional `migration' between them. This iterative process of fitness evaluation and mutation continues across multiple generations until a satisfactory outcome is attained or until a user-specified condition is satisfied. The decision to utilise PySR was based on its highly customizable algorithm, which permits modifications in loss functions, complexity constraints, and allows customisation of the functional form of the final expressions. This code has been used previously in many scientific studies \citep{sr_example2, sr_example3, sr_example1}.

Our training data consist of the dewiggled linear and nonlinear matter power spectra (see Section \ref{sec:methodology_bao}) and the set of cosmological parameters $\theta \; \mathrm{or} \; \theta_X$ based on the model (see Section \ref{sec:methodology_power_spectrum}). PySR provides the capability to identify expressions conforming to a user-defined functional form. This feature enhances computational efficiency by constraining the expression search space to only those equations that adhere to the specified format. We utilise this capability by fitting to a specific function of the shape given by equation~\eqref{eqn:func_form_SR}.

We let our model evolve for 1000 generations with a log loss. As we are finding a function for $\Delta^2_{\mathrm{NL}}(k_{\mathrm{NL}})$, applying a log loss ensures that the residuals are small even for extremely small values of the dimensionless power spectrum. Our function consists of basic operators $+,\;-,\;\times\;\text{and}\;\div$. SR is often characterised as a Pareto optimisation problem \citep{exhaustive_sr}, wherein the objective is to identify a solution that achieves an adequate level of accuracy while simultaneously minimising complexity. Within the framework of PySR, complexity is quantified by the number of nodes present in the expression tree corresponding to the equation. A Pareto curve is a plot of the loss versus the complexity of the equation, and aids in identifying potential solutions.  

Unlike traditional ML algorithms, genetic algorithms tend not to converge to a singular solution. Rather, they have the capability to identify a spectrum of new equations and to explore this newly defined space for potential solutions endlessly. In this context, the Pareto curve is of paramount importance, as it indicates when a viable expression has been produced. The most optimal equations reside on the Pareto front. This curve contains the best solutions found by SR at each particular complexity value. To determine the superior model, it is crucial to concentrate exclusively on the models that reside along this front, as they represent the most effective solutions to the problem being addressed. Moreover, an important attribute of Pareto curves is their tendency to reach a plateau after a certain level of complexity is attained. This phenomenon suggests that increasing complexity may lead to diminishing returns in accuracy. 

Alongside the Pareto front, we quantify the goodness of fit for our analytical function by plotting residuals of our model compared to Quijote and also by measuring the mean absolute percentage error (MAPE), which is defined as
\begin{equation}
    \mathrm{MAPE}\left(P_{\mathrm{Quijote}}, P_\mathrm{Analytical}\right) = \frac{1}{N} \sum_{i=0}^{N-1} \frac{\left | P_\mathrm{i,Quijote} - P_\mathrm{i,Analytical}\right|}{\mathrm{max}\left(\epsilon, \left|P_\mathrm{i,Quijote}\right| \right)},
\end{equation}

\noindent where $\epsilon$ is a strictly positive number, such that we avoid undefined results. This number is arbitrarily small. In this study, we use the MAPE function provided by \cite{scikit-learn}.

\subsection{Calculating $P_{\mathrm{NL}}(k_{\mathrm{NL}})$} \label{sec:methodology_calculate_pk}
The method to go from an input linear power spectrum to the final prediction of the nonlinear power spectrum, $P_{\mathrm{NL}}(k_{\mathrm{NL}})$, is fairly straightforward. First, we obtain the prediction for the nonlinear dimensionless power spectrum $\Delta_{\mathrm{NL}}^2(k_{\mathrm{NL}})$ for our chosen model as a function of cosmological parameters by substituting the nonlinear correction term $N\left[\Delta_{\mathrm{L}}^2(k_{\mathrm{L}}),\left(\theta \;\text{or}\; \theta_X\right) \right]$ in equation~\eqref{eqn:func_form_SR}. We can then convert this dimensionless power spectrum prediction to $P_{\mathrm{NL}}^{\mathrm{nw}}(k_{\mathrm{NL}})$ using equation~\eqref{eqn:pk_conversion}. Since this is the `no-wiggle' version of the nonlinear power spectrum, we reintegrate the BAOs by rearranging the terms in equation~\eqref{eqn:nl_dewiggle} to obtain the final prediction for the nonlinear power spectrum $P_{\mathrm{NL}}(k_{\mathrm{NL}})$.

In the upcoming sections, we present the method behind choosing the best equation, the best fit equations and fits for each model. Finally, we talk about the supplemental equations PySR generates.

\section{Results} \label{sec:results}
In this section, we begin by describing the method through which PySR selects the best equation from all the equations contained within the Pareto front. This is followed by the equations that make up \spine and \spinex. Subsequently, followed by the $P_{\mathrm{NL}}(k_{\mathrm{NL}})$ fits for Planck-like, fiducial and LH cosmologies, respectively. Finally, we discuss the supplementary equations produced by SR.

\subsection{Selecting the best equation} \label{sec:results_select_eqn}
PySR selects the best equation based on the value of the model selection hyperparameter. This determines the best equation at each complexity. This hyperparameter can have three possible values.
\begin{enumerate}
    \item `accuracy': This selects a model with the lowest loss, i.e., the most accurate equation. However, it is important to note that this pursuit of accuracy may necessitate a compromise in interpretability, as the most precise models are often characterised by greater complexity.
    \item `best': This selects a model with the largest score within the group of equations that have a loss exceeding 1.5 times the most accurate model.
    \item `score': This selects an expression with the highest score. In PySR, score is defined as the negative derivative of the log-loss with respect to complexity\footnote{\url{https://ai.damtp.cam.ac.uk/pysr/api/}}. 
\end{enumerate}

We set this hyperparameter to `best' as we considered it a suitable compromise between `accuracy' and `score'. This selection reflects our objective of choosing an equation that offers sufficient accuracy while avoiding unnecessary complexity. Importantly, `best' is also the default value for this parameter in PySR. Furthermore, we opted to enhance the complexity of the equation, which is regulated by the `maxsize' hyperparameter. We determined a value of 80, in contrast to the default value of 30. This decision was made to ensure an adequate fit for the power spectrum, as our findings indicated that equations with complexities of 30 and below were insufficient for this purpose. As a result of this combination of hyperparameters, SR generated 39 candidate solutions for \spine and 35 solutions for \spinex with complexities ranging from 1 to 79 for both models.

\begin{figure}
    \centering
    \includegraphics[width=\columnwidth]{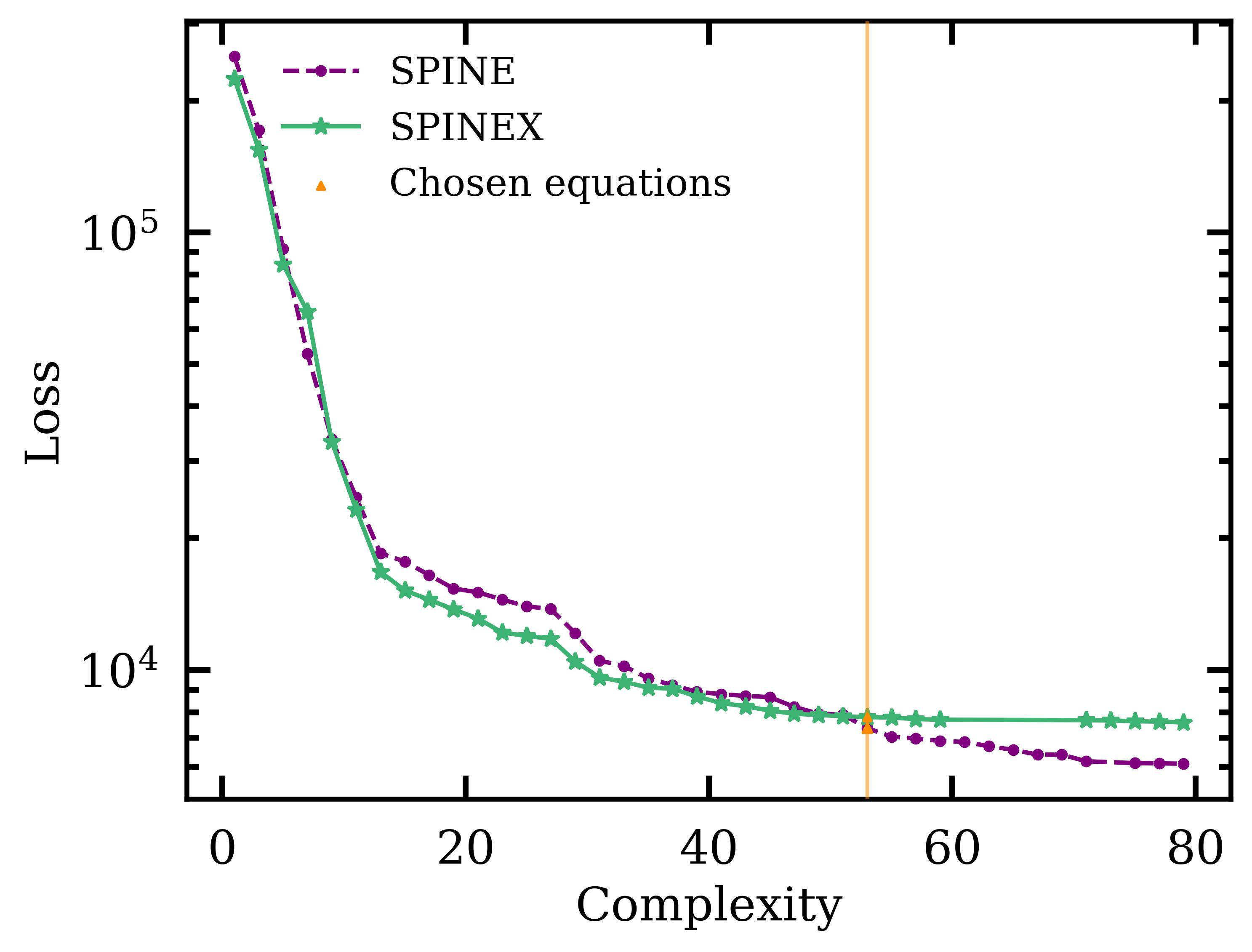}
    \caption{Pareto front for \spine (purple dashed line) and \spinex (green solid line). The points and stars illustrate the equations for that level of complexity in each model. Selected equations have been distinctly marked with a yellow triangle and a vertical line to enhance clarity. The equations exhibit a complexity level of 53 for both \spine and \spinex.}
    \label{fig:pareto_curve}
\end{figure}

These possible solutions are represented by the two Pareto fronts of Fig.\ref{fig:pareto_curve}, in which we have marked the best solutions for each respective model in yellow. This figure also exhibits several notable features. There are several dips in the loss as the complexity of the equations increases. This can be attributed to the fact that, as SR explores the search space for new equations, it uncovers new combinations of variables and operators that effectively align with the data at that level of complexity. 

The large decrease starting at complexity 1 and ending at complexity 9 in Fig.\ref{fig:pareto_curve}, for both the \spine and \spinex curves, is anticipated, as functions belonging to these complexities are simple linear combinations of the input variables and are not expected to fit the power spectrum with much accuracy. Equations become more accurate as their complexity increases, until we reach the start of the Pareto plateau, where any further increase in complexity does not result in an improvement in the fit. 
For example, the best solutions for \spine and \spinex both have a complexity of 53 and lie on the Pareto front. We choose these expressions for the emulators. The Pareto fronts for the two models behave similarly, with any real deviation occurring beyond complexity 53. It is essential to highlight that the model selected for \spinex was specifically chosen through visual inspection to maximise accuracy. To allow for a fairer comparison between \spine and \spinex approaches, we use the same complexity for both models.

\begin{table}
\caption{Summary of the Mean Absolute Percentage Error (MAPE) for \spine, \spinex, and alternate power spectrum emulator codes mentioned in this paper. Note here that models whose parameter ranges do not encompass the cosmologies examined in this study are shown with dashes. We would also like to comment here that SYREN-Halofit also extrapolates beyond its parameter range for LH cosmologies.}
    \centering
    \begin{tabular}{lcccc}
    \hline
    \textbf{Model} & \multicolumn{3}{c}{\textbf{MAPE [\%]}} \\
               & \textbf{LH} & \textbf{Fiducial} & \textbf{Planck like} \\
    \hline
        SPINE           & \LHmapeSPINE  & \FIDmapeSPINE  & \PLANCKmapeSPINE  \\
        SPINEX          & \LHmapeSPINEX & \FIDmapeSPINEX & \PLANCKmapeSPINEX \\
        Halofit       & 5.24          & 1.37           & 3.08              \\
        HMCode          & 4.89          & 2.83           & 3.97              \\
        EuclidEmulator2 & $-$           & 0.45           & $-$               \\
        Aletheia        & $-$           & 0.87           & $-$               \\
        SYREN-Halofit   & 5.86          & 1.08           & 1.96              \\
    \hline
    \end{tabular}
    \label{tab:mape_models}
\end{table}

In the upcoming sections, we present the functional forms of our \spine and \spinex recipes for the nonlinear power spectrum $P_{\mathrm{NL}}(k_{\mathrm{NL}})$ as a function of the cosmological parameters described in Section \ref{sec:methodology_power_spectrum}. We will depict the fits for all the LH cosmologies that lie within 20$\sigma$ of the Planck-2018 observations, the fiducial cosmologies utilised in Quijote, and fits for the LH cosmologies, respectively.

\subsection{Fits for \spine and \spinex} \label{sec:results_spine_eqn}

The nonlinear correction term obtained through SR for \spine is given as
\begin{multline}\label{eqn:spine_final_eqn}
N\left[\Delta_{\mathrm{L}}^2(k_{\mathrm{L}}),\theta \right] 
= \left(\Delta_{\mathrm{L}}^2(k_{\mathrm{L}}) \right)^{2.64}
\Bigl[\Delta_{\mathrm{L}}^2(k_{\mathrm{L}})
\Bigl(
\left( 4.16^{n_{\mathrm{L}}} g_{\mathrm{a}}^{n_{\mathrm{s}}}\,\Pi \right)^{n_{\mathrm{L}}} \\
- 0.16
\Bigr)
- n_{\mathrm{L}} \Bigr].
\end{multline}

\noindent Where $n_\mathrm{L}$ is the slope of the late-time linear power spectrum (see Table \ref{tab:model_params}). Here,
\begin{equation}
    \Pi = \left[
    -n_\mathrm{L} + \frac{(9.13^{n_{\mathrm{L}}}n_{\mathrm{s}})^{\Delta_{\mathrm{L}}^2(k_{\mathrm{L}})} \left( 162.45 + \alpha \right)}{\sigma_8 g_{\mathrm{a}}} \right]^{\Delta_{\mathrm{L}}^2(k_{\mathrm{L}})} + 4.24
\end{equation}
and
\begin{equation}
\alpha = \frac{n_{\mathrm{L}}^3 + 3.76/\Delta_{\mathrm{L}}^2(k_{\mathrm{L}})}{\Omega_{\mathrm{m}}-\Omega_{\mathrm{b}}}.
\end{equation}
The nonlinear correction term for \spinex is 
\begin{equation}
\label{eqn:spinex_final_eqn}
    N\left[\Delta_{\mathrm{L}}^2(k_{\mathrm{L}}),\theta_X \right] = \frac{\left(\Delta_{\mathrm{L}}^2(k_{\mathrm{L}})\right)^{4.305}}{n_s} \left(0.085\Delta^2_{\mathrm{L}}(k_{\mathrm{L}})(\kappa_1 + \kappa_2) + \varepsilon\right)
\end{equation}
where
\begin{align}
    \kappa_1 &= \left(\left(\left(n_\mathrm{s}^{\tilde{x}} + f_\mathrm{b}^{\mathrm{-0.154}} - 0.652\tilde{x}\Delta^2_{\mathrm{L}}(k_{\mathrm{L}})\right)^{\sigma_{\mathrm{12}}} + \tilde{x} + f_\mathrm{b}\right)\tilde{x} \right)^{n_\mathrm{L}^2},\\
    \kappa_2 &= \frac{f_{\mathrm{b}}-\tilde{x}^{-15.169}}{\sigma_{12}} 
\end{align} 
and 
\begin{equation}
   \varepsilon = \frac{\tilde{x}+f_{\mathrm{b}}-0.339}{\Delta_{\mathrm{L}}^2(k_{\mathrm{L}})}.
\end{equation}
\noindent In the next three sections, we present the $P_{\mathrm{\mathrm{NL}}}(k_{\mathrm{NL}})$ predictions for Planck-like, fiducial and LH cosmologies, respectively, for both models.

\subsubsection{Cosmologies within $20\sigma$ of Planck-2018} \label{sec:results_planck}
By constraining the parameter space to be within 20$\sigma$ of the Planck-2018 best-fit parameters, we can ensure that our predictions remain observationally relevant and statistically meaningful. Moreover, this restriction illustrates the performance of our emulator in regions that are most likely to describe our Universe, while maintaining a connection to the fiducial cosmology around which many forecasts and analyses are performed. We have restricted the LH space in such a way that each of the five cosmological parameters (see Section \ref{sec:data}) remains within 20$\sigma$ of Planck individually.

\begin{figure*}
\centering
    \begin{subfigure}[t]{0.48\textwidth}
        \vspace{0.5cm}
        \includegraphics[width=\columnwidth]{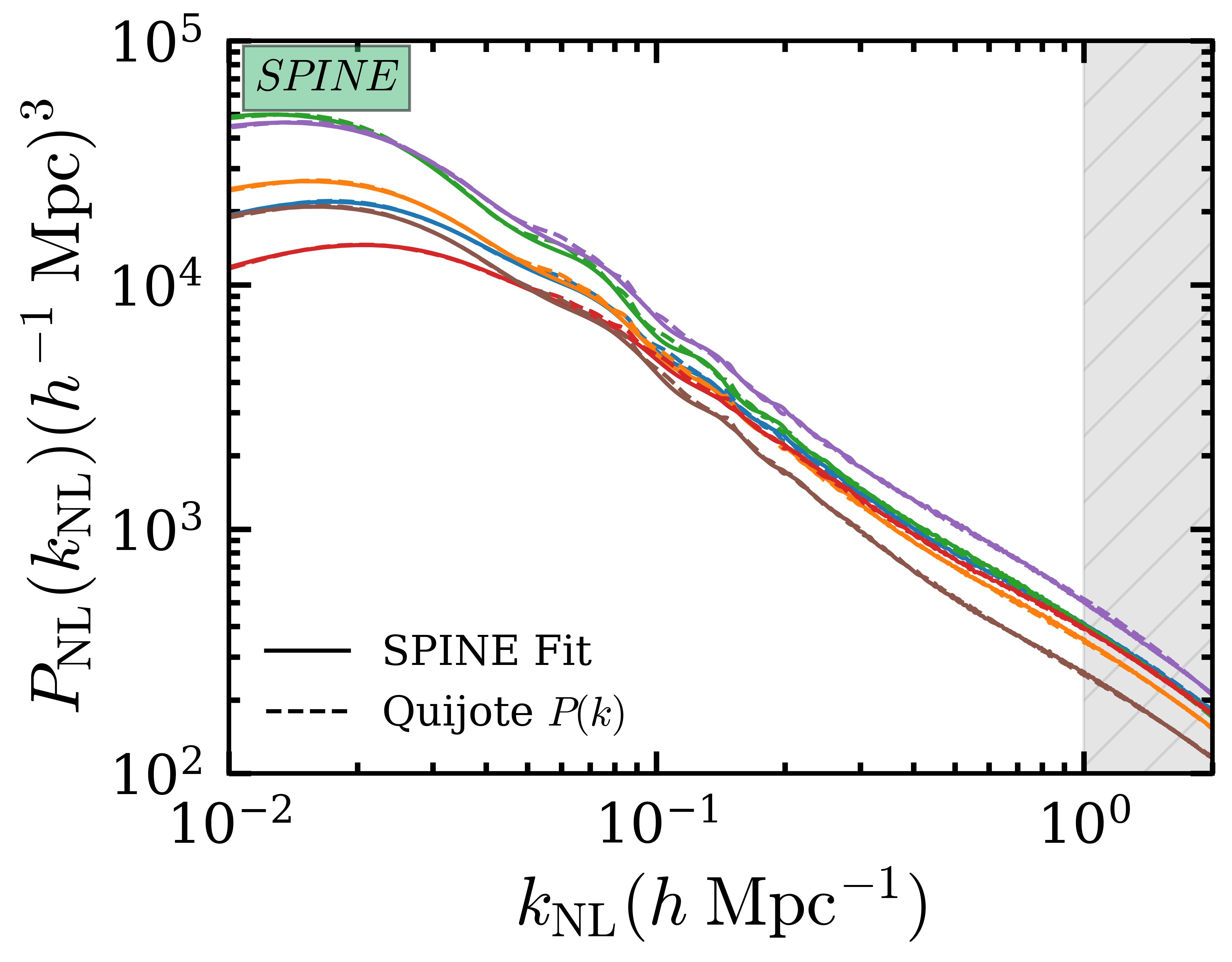}
        \caption{}
        \label{subfig:planck_spine_pk}
    \end{subfigure}
\hfill
    \begin{subfigure}[t]{0.5\textwidth}
        \vspace{0pt}
        \includegraphics[width=\columnwidth]{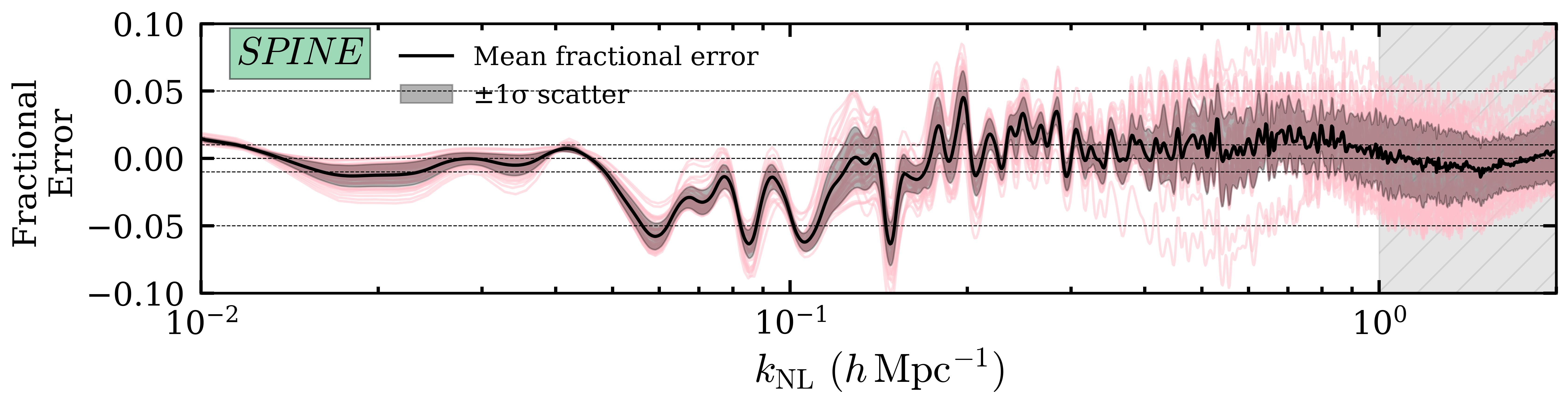}
        \caption{}
        \label{subfig:planck_spine_residuals}
        \vspace{0.05cm}

        \includegraphics[width=\columnwidth]{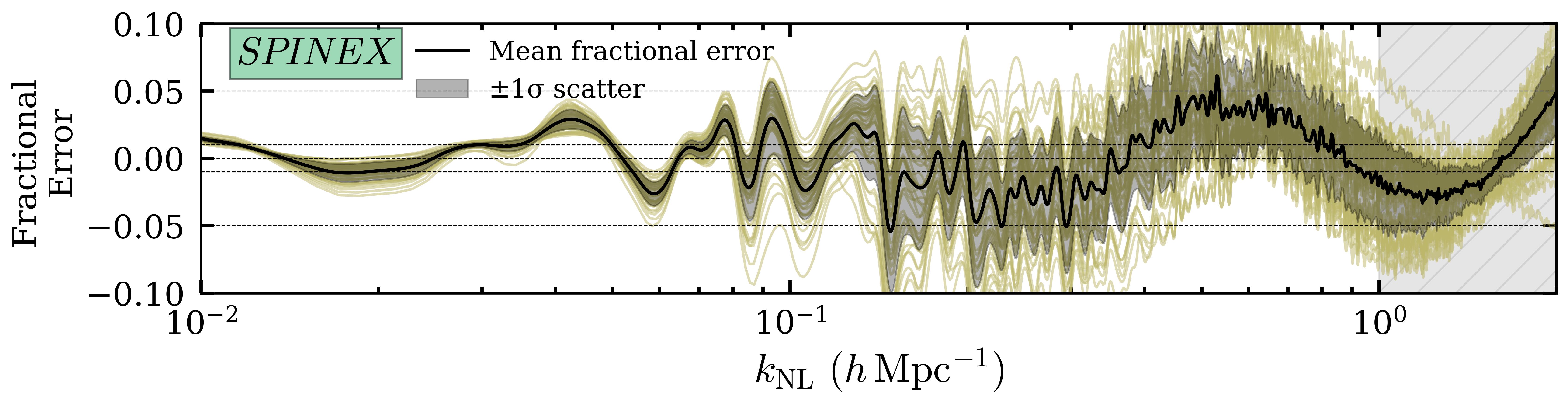}
        \caption{}
        \label{subfig:planck_spinex_residuals}
        \vspace{0.05cm}

        \includegraphics[width=\linewidth]{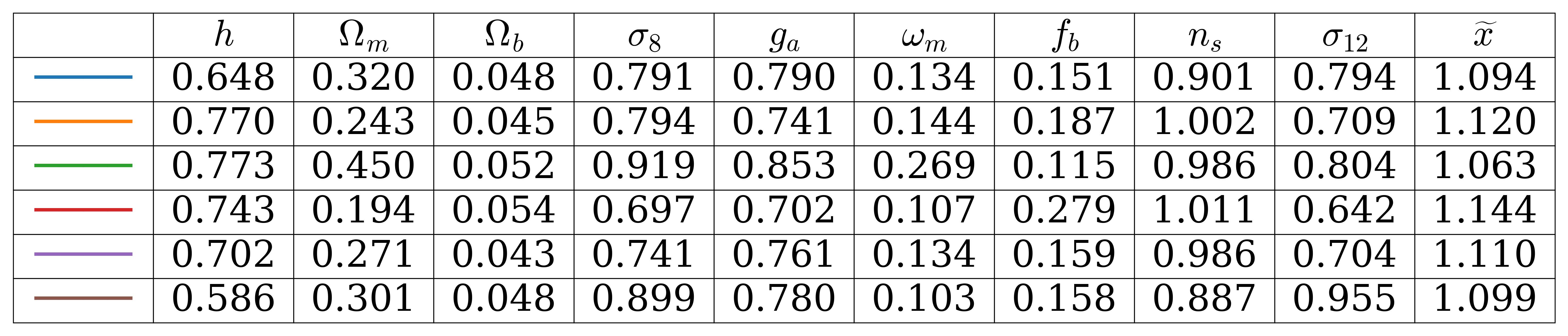}
        \caption{}
        \label{subfig:planck_random_cosmos}
    \end{subfigure}

\caption{\textbf{(a)}: Fits for the nonlinear power spectrum obtained using \spine (solid lines) compared to the Quijote nonlinear power spectrum (dashed lines) for a random subset of six cosmologies from the 45 cosmologies that lie within 20$\sigma$ of Planck-2018. Panel \textbf{(d)} lists the cosmological parameters for this random subset. Panels \textbf{(b)} and \textbf{(c)} depict the fractional error for all 45 cosmologies for \spine and \spinex, respectively. Additionally, we show the mean of fractional error (black solid line) and the 1$\sigma$ region (black shaded region) for each respective model. The black dashed lines represent the 1\% and 5\% margins. The grey hatched regions in (a), (b) and (c) depict the region where the Quijote simulations have converged at $>2.5\%$.}
\end{figure*}

Within the 2000 cosmologies of the Quijote suite, 45 cosmologies are within the 20$\sigma$ threshold. For visualisation purposes, we randomly selected six cosmologies from this set and compared them against our model predictions. For \spine, these cosmologies are plotted in Fig.\ref{subfig:planck_spine_pk}, and the parameters associated with these spectra are listed in the table in Fig.\ref{subfig:planck_random_cosmos}. We have plotted all 45 Planck-like cosmologies in Fig.\ref{subfig:planck_spine_residuals}. The black curve depicts the mean of fractional error of these cosmologies, with the black shaded region illustrating the 1$\sigma$ region around the mean. These 45 cosmologies have a MAPE of \PLANCKmapeSPINE\%. For \spinex, the same cosmologies have been plotted in Fig.\ref{subfig:planck_spinex_residuals} with the mean and variance. \spinex achieves a MAPE of \PLANCKmapeSPINEX\% for the Planck-like cosmologies, indicating a high level of accuracy across all scales analysed.

For these random cosmologies, the power spectrum predictions are accurate on a wide range of scales. One can also see large `wiggles' corresponding to $k\sim0.1 \;h \;\mathrm{Mpc}^{-1}$. These are not due to the BAO signal but are rather caused by cosmic variance in the simulation, and we discuss this in detail in Section \ref{sec:discussion_model_performance}. Regardless of this, both models provide an excellent fit for all the $k$-scales that we have considered. Table \ref{tab:mape_models} contains the MAPE values for Planck-like cosmologies, as computed by \spine, \spinex and other models examined in this work.

\subsubsection{Fiducial cosmology} \label{sec:results_fiducial}
Fiducial cosmologies are often chosen to be close to current observational data (e.g. Planck-2018 $\Lambda$CDM results), and making predictions at these cosmologies is most directly useful for comparison with real data. We use the prediction from \spine and \spinex to test whether the interpolation from the LH space is robust at the Quijote fiducial cosmology. Additionally, anchoring predictions at the fiducial cosmology facilitates the comparison of additional emulators at a common reference. Since numerous cosmological analyses are centred around a fiducial model, obtaining fast and accurate predictions for the fiducial model is particularly important. We therefore consider fits for the Quijote fiducial cosmology, with the parameter values detailed in Table.\ref{tab:Quijote_limits}. 

To obtain a prediction for the fiducial power spectrum, we take the mean of the 500 paired fixed fiducial simulations \citep[see][for more details]{stats_paired_fixed}. The averaged power spectrum and the corresponding fiducial cosmological parameters are then used to obtain a prediction for $P_{\mathrm{NL}}(k_{\mathrm{NL}})$ using both symbolic models.

\begin{figure}
    \centering
    \includegraphics[width=\columnwidth]{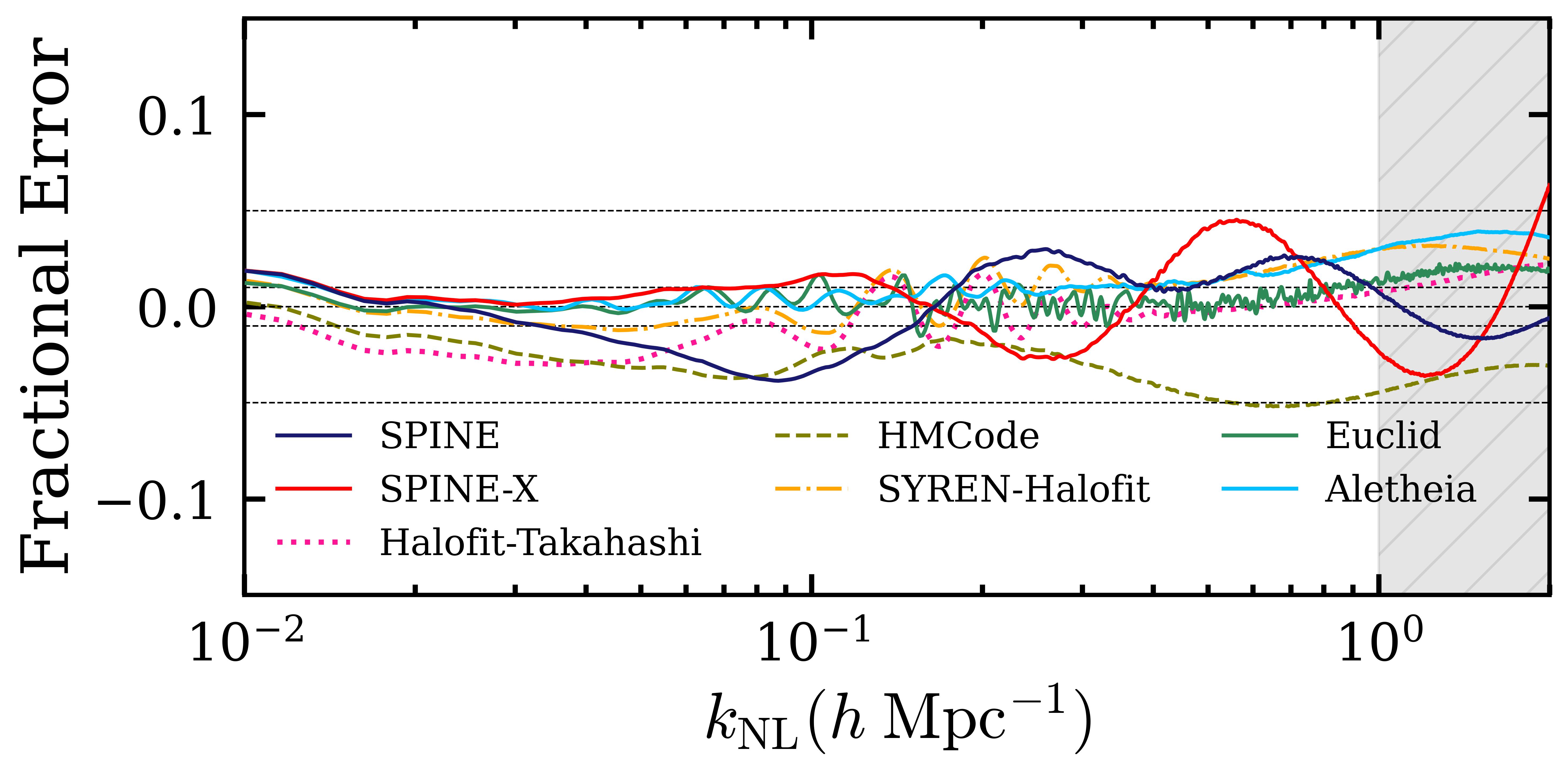}
    \caption{Power spectrum prediction calculated by \spine (navy solid line) and \spinex (red solid line) compared to other emulators considered here, namely Halofit-Takahashi (pink dotted line), HMCode (olive dash-dotted line), EuclidEmulator (green solid line), Aletheia (light blue solid line) and SYREN-Halofit (yellow dash-dotted line) as plotted for the ensemble average of the Quijote fiducial cosmologies. The dashed lines are the 5\% and 1\% error margins. The grey hatched region indicates the region where the Quijote simulations have converged at $>2.5\%$.}
    \label{fig:fiducial_compare}
\end{figure}

We show the \spine and \spinex residuals compared to Quijote in Fig.\ref{fig:fiducial_compare}. For the fiducial cosmology, the fit is in line with other emulators considered here, namely Halofit (Takahashi model) \citep{halofit_takhashi}, HMCode \citep{hmcode}, Euclid Emulator \citep{euclid_emulator,euclid_emulator2}, Aletheia \citep{aletheia} and SYREN-Halofit \citep{syren-halofit}, another SR-enabled symbolic emulator. \spine has a MAPE of \FIDmapeSPINE\% for the fiducial cosmology while \spinex exhibits a MAPE of \FIDmapeSPINEX\%. The two models remain consistent with the other emulators considered here and stay within 5\% for a majority of $k$-scales. Table \ref{tab:mape_models} consists of the MAPE values associated with fiducial cosmology for other models. Note here that \spine and \spinex have been trained to replicate the Quijote simulations $P_{\mathrm{NL}}(k_{\mathrm{NL}})$, which have low resolution. In contrast, other emulators such as EuclidEmulator2 and Aletheia provide predictions corresponding to higher resolution simulations. This discrepancy will affect the performance of the models in the current comparison. Despite this, the \spine and \spinex produce outputs with comparable accuracy. Additionally, \spinex shows oscillatory behaviour around $k_\mathrm{NL}\sim0.5\;h\;\mathrm{Mpc}^{-1}$ in the fiducial and Planck-like case. This behaviour is attributed to \spinex's limitations in accurately capturing the shape of nonlinear structures rather than oscillatory terms in the expressions themselves.

\subsubsection{LH cosmologies} \label{sec:results_LH}
\begin{figure*}
\centering
    \begin{subfigure}[t]{0.48\textwidth}
        \vspace{0.5cm}
        \includegraphics[width=\columnwidth]{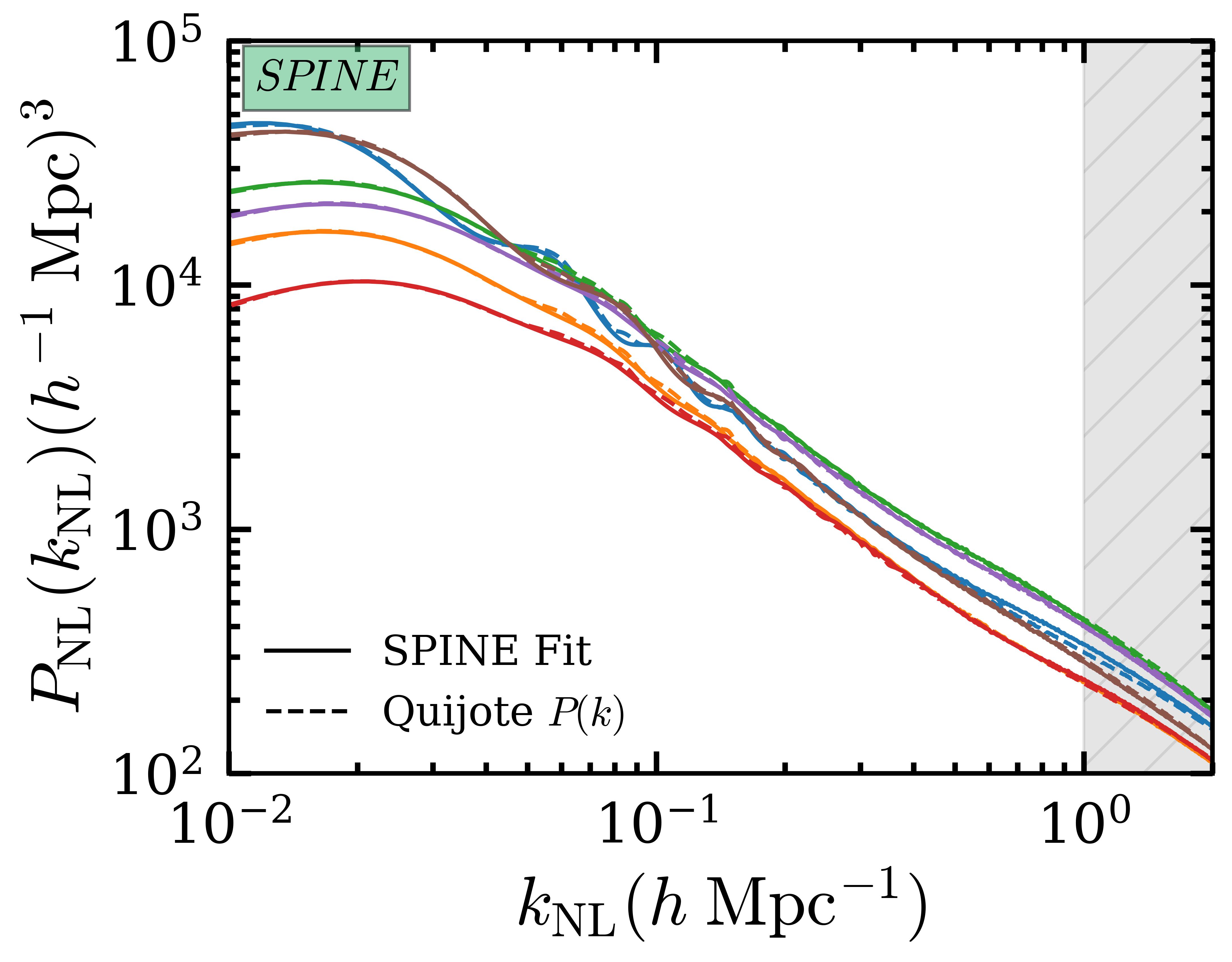}
        \caption{}
        \label{subfig:LH_spine_pk}
    \end{subfigure}
\hfill
    \begin{subfigure}[t]{0.5\textwidth}
        \vspace{0pt}
        \includegraphics[width=\columnwidth]{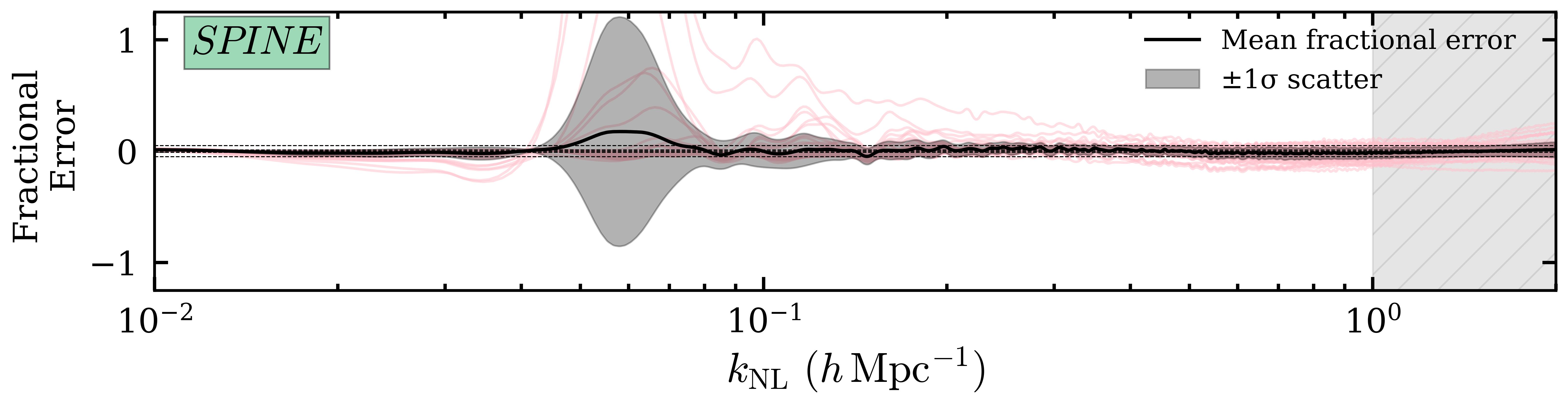}
        \caption{}
        \label{subfig:LH_spine_residuals}
        \vspace{0.05cm}

        \includegraphics[width=\columnwidth]{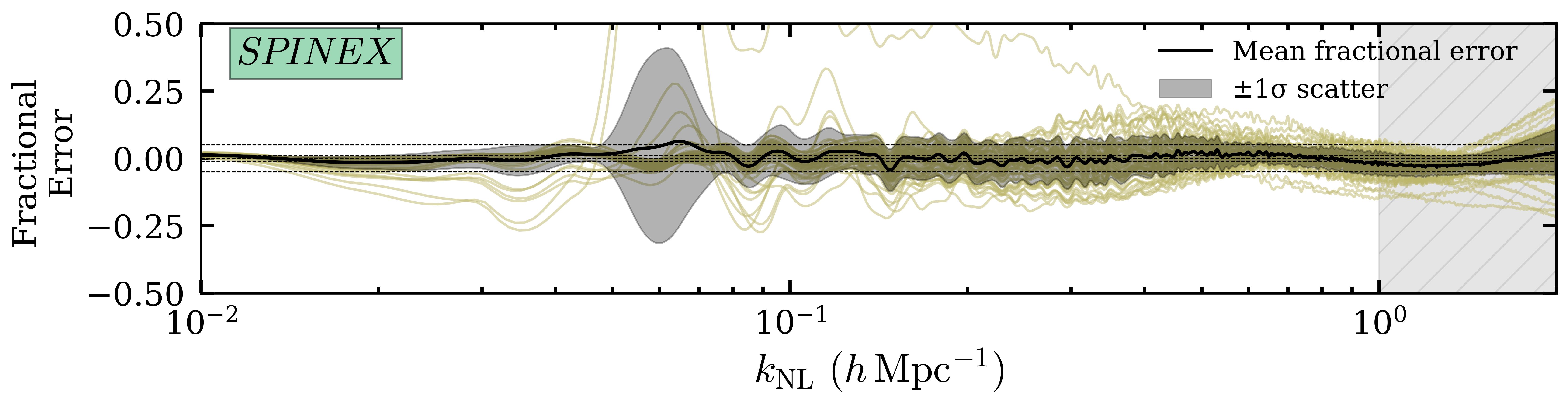}
        \caption{}
        \label{subfig:LH_spinex_residuals}
        \vspace{0.05cm}

        \includegraphics[width=\linewidth]{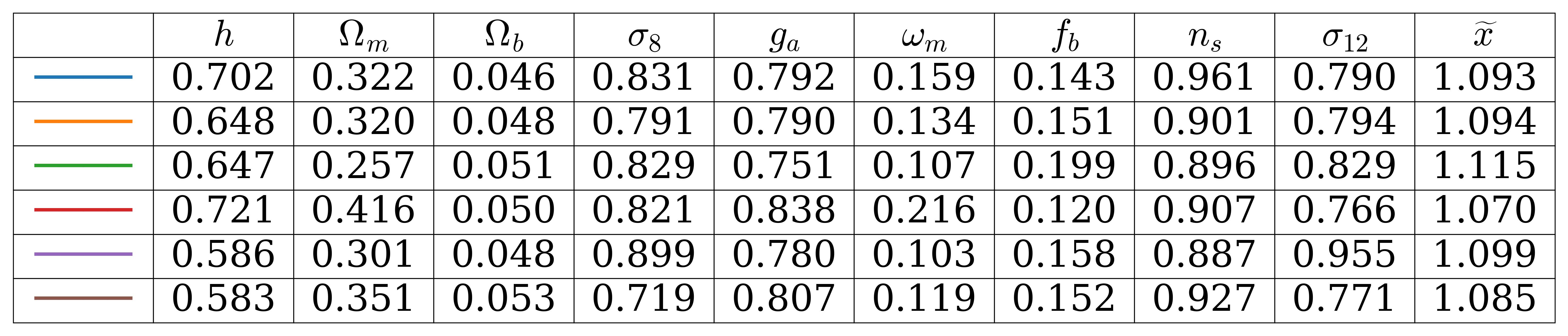}
        \caption{}
        \label{subfig:LH_random_cosmos}
    \end{subfigure}

\caption{\textbf{(a)}: Example fits for the nonlinear power spectrum computed using \spine (solid lines) compared to the Quijote nonlinear power spectrum (dashed lines) for a subset of six random cosmologies from the 68 test cosmologies. Panel \textbf{(d)} lists the cosmological parameters for the random subset. Panels \textbf{(b)} and \textbf{(c)} illustrate the fractional error for all cosmologies in the test set for \spine and \spinex, respectively. We also plot the mean of the fractional error (black solid line) and the 1$\sigma$ region (black shaded region) for each respective model. The dashed lines represent the 1\% and 5\% margins. The grey hatched regions depict the region where the Quijote simulations have converged at $> 2.5\%$.}
\end{figure*}

For visual inspection, we have presented the $P_{\mathrm{NL}}(k_{\mathrm{NL}})$ prediction for a subset of six random cosmologies from the test set. Fig.\ref{subfig:LH_spine_pk} shows the \spine prediction for these cosmologies. The parameters associated with these cosmologies can be found in the table in Fig.\ref{subfig:LH_random_cosmos}. Figs. \ref{subfig:LH_spine_residuals} and \ref{subfig:LH_spinex_residuals} depict the residuals compared to Quijote for all 68 test cosmologies for \spine and \spinex respectively. In these figures, we have plotted the mean of the fractional error and its variance for these cosmologies. \spine achieves a MAPE of \LHmapeSPINE\% and \spinex has a MAPE of \LHmapeSPINEX\%.  

As mentioned previously, the wiggles around the value $k\sim 10^{-1}\;h\;\mathrm{Mpc}^{-1}$ arise from cosmic variance. Regardless, both models provide a good description of the power spectrum across a wide range of $k$-scales analysed. We direct the reader to Table \ref{tab:mape_models} for a comparative analysis of the MAPE values associated with the LH cosmologies, as calculated through the various models examined in this work.

\subsection{Auxiliary Equations from PySR} \label{sec:results_auxilliary_eqns}
\begin{figure}
    \centering
    \includegraphics[width=1\columnwidth]{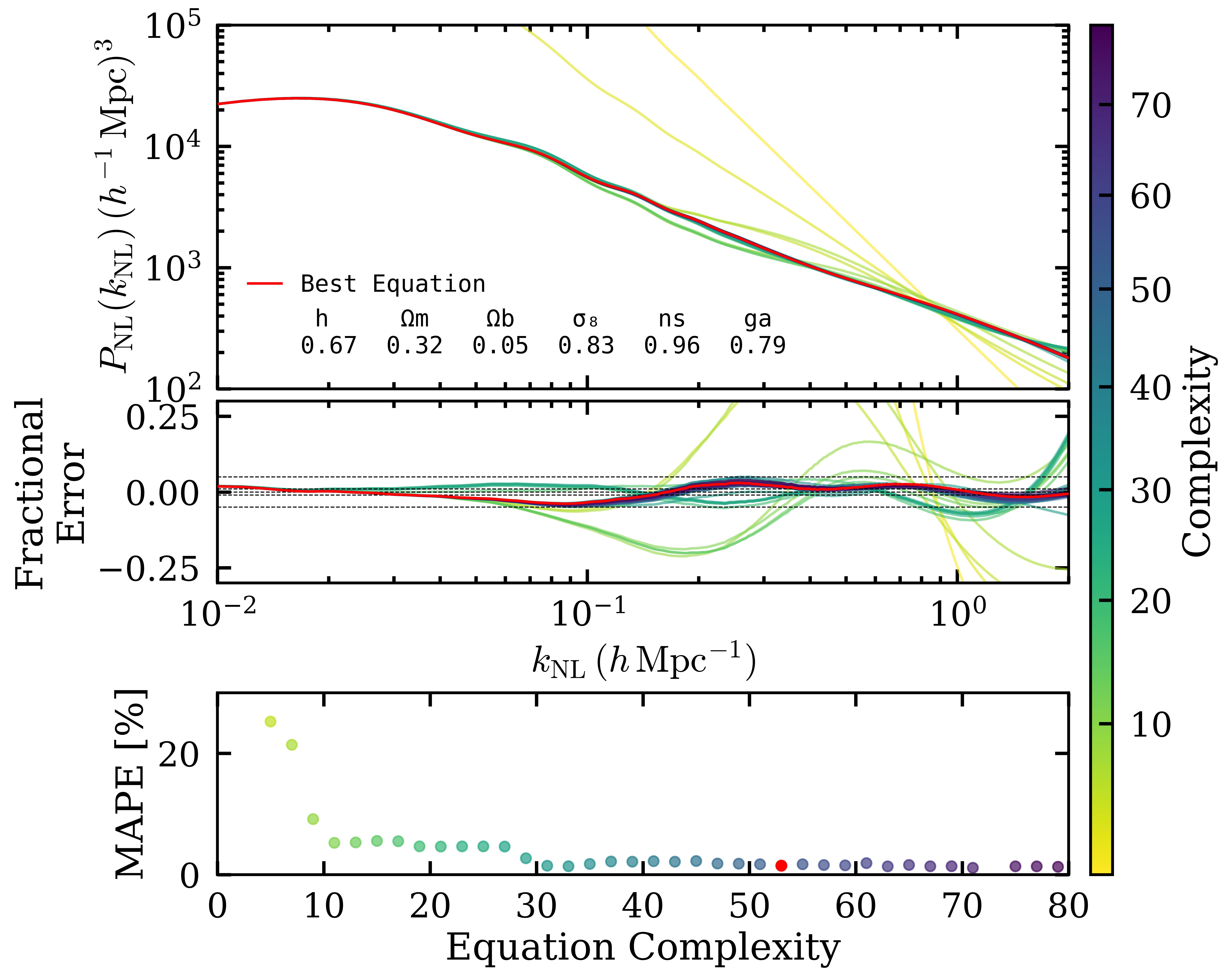}
    \caption{All the equations generated from the final SR iteration predicting the fiducial power spectrum, coloured by their complexity value. The best solution, outlined in Section \ref{sec:results} (given in equation~\ref{eqn:spine_final_eqn}), is plotted in red. \textbf{Top}: Fiducial power spectrum prediction for each equation generated by SR. \textbf{Middle}: Residuals compared to the averaged Quijote fiducial cosmology. The dotted lines represent the 5\% and 1\% thresholds. \textbf{Bottom}: MAPE for each cosmology as a function of complexity.}
    \label{fig:all_eqn_plot}
\end{figure}

In this section, we aim to understand and reinforce the reason for PySR's choice of equation. For this task, we will use the PySR output for \spine (equation~\ref{eqn:spine_final_eqn}). In Fig.\ref{fig:all_eqn_plot}, we plot all potential solutions for the fiducial cosmology spawned at each complexity. The chosen solution is marked in red in all plots.  

The analysis of the power spectrum predictions (Fig.\ref{fig:all_eqn_plot} top panel), and their residuals with respect to Quijote (Fig.\ref{fig:all_eqn_plot} middle panel), demonstrates that expressions exhibiting complexities $\leqslant$15 are not considered viable candidates for this study, as they fail to fit the power spectrum adequately. Conversely, it is observed that expressions with complexities ranging from 16 to 25 tend to offer slightly better accuracy at the BAO scale, although this is at the expense of significant degradation in accuracy at smaller scales, with deviations exceeding the 10\% level for the smallest scales considered here. With increasing complexity, the agreement improves, achieving optimal results around complexity 53. Beyond this threshold, further increases in complexity do not yield substantial enhancements in the fit. This conclusion is further substantiated by the MAPE for each expression plotted as a function of the equations' complexity (Fig.\ref{fig:all_eqn_plot} bottom panel) and corroborated by the green line in the Pareto front illustrated in Fig.\ref{fig:pareto_curve}. These two graphical representations can be regarded as illustrating equivalent information. Overall, the best solution chosen by SR in this case has a sub-5\% accuracy for a majority of scales considered in this study and forms the basis of \spine (see equation~\ref{eqn:spine_final_eqn}). The same analysis can be done for \spinex. However, in this case, the best solution was chosen by visual inspection of all potential symbolic solutions. The chosen equation for \spinex has a complexity of 53 and performs with a sub-5\% accuracy for all $k$-scales analysed (see equation~\ref{eqn:spinex_final_eqn}). 

\section{Discussion} \label{sec:discussion}
In this section, we aim to facilitate a clearer understanding of how several key parameters influence the power spectrum, thereby providing an intuitive underpinning for both models.

\subsection{Comparative analysis of \spine and \spinex} \label{sec:discussion_understanding}
\begin{figure}
    \centering
    \includegraphics[width=\columnwidth]{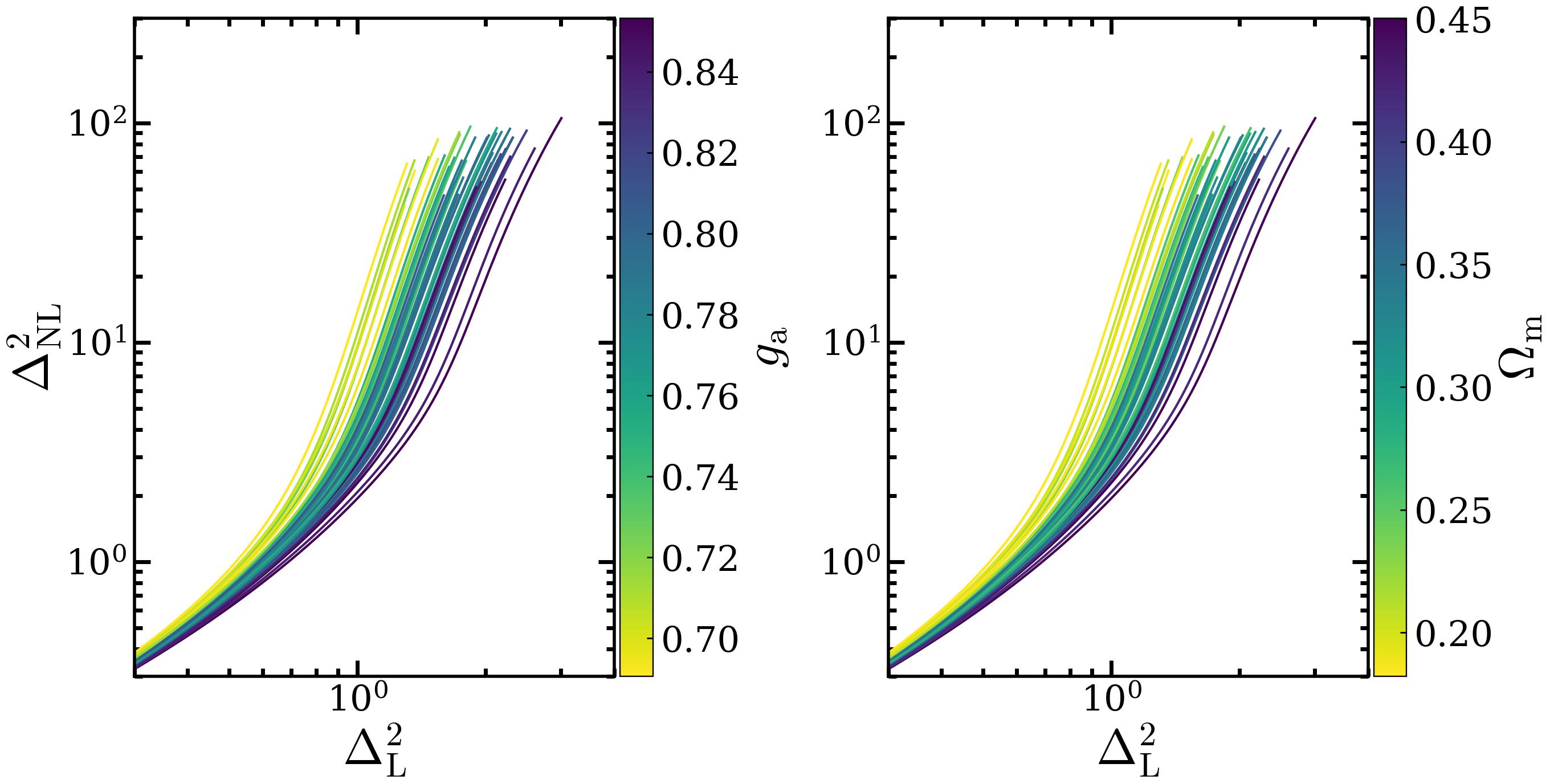}
    \caption{Dimensionless power spectrum coloured according to their value of the growth suppression factor $g_\mathrm{a}$ and matter density $\Omega_\mathrm{m}$ respectively. All of the plots are for cosmologies within 20$\sigma$ of Planck as produced by \spine.}
    \label{fig:how_params_change_spine_mainpaper}
\end{figure}

Fig.~\ref{fig:how_params_change_spine_mainpaper} illustrates the \spine dimensionless power spectrum as plotted for the cosmologies that are within 20$\sigma$ of Planck-2018. This functions as a reference for identifying the onset of nonlinearities and highlights the most important parameters in this context. The growth suppression factor $g_{\mathrm{a}}$ is the only ingredient essential in guiding the advent of nonlinear evolution. It determines the relative suppression of structure compared to an Einstein-de Sitter universe. If $g_{\mathrm{a}}$ is smaller, perturbations take longer to reach the nonlinear threshold. This results in the shift of the nonlinear scale (where $\Delta_{\mathrm{NL}}^2(k_{\mathrm{NL}}) \sim 1$) to smaller values at a given epoch. The growth suppression factor sets the time evolution of the normalisation of clustering, i.e. two cosmologies can have the same initial spectrum, but if they are characterised by different $g_{\mathrm{a}}$ they will yield different $P_{\mathrm{NL}}(k_{\mathrm{NL}})$ simply because they have different rates of nonlinear structure formation \citep{evolution_mapping, aletheia}. 

In addition to this, Fig.\ref{fig:how_params_change_spine_mainpaper} also shows that the non-linear power spectrum has a dependency on $\Omega_{\mathrm{m}}$. However, $\Omega_{\mathrm{m}}$ and $g_a$ are closely related, as  
\begin{equation}
    g(\Omega) = \frac{5}{2}\left[\Omega_{\mathrm{m}}^{4/7} - \Omega_{\mathrm{v}}+\left(1+\frac{\Omega_{\mathrm{m}}}{2} \right)\left(1+\frac{\Omega_{\mathrm{v}}}{70} \right) \right]^{-1},
\end{equation}
where $\Omega_{\mathrm{m}}$ and $\Omega_{\mathrm{v}}$ are the matter and vacuum contributions respectively. This suggests that $g_{\mathrm{a}}$ is the main quantity required to account for the cosmology dependence of the solutions. This was also a prediction of PD96, where the cosmology dependence only centres through the factor $g_{\mathrm{a}}$, and it governs the amplitude of the virialised portion of the spectrum. However, contrary to \cite{jain_mo_white}, we do not see any significant dependence on $n_{\mathrm{s}}$, although this could be because we do not have any values which are $n_\mathrm{s}\lesssim -1$, where this dependency is most pronounced.

\begin{figure}
    \centering
    \includegraphics[width=\columnwidth]{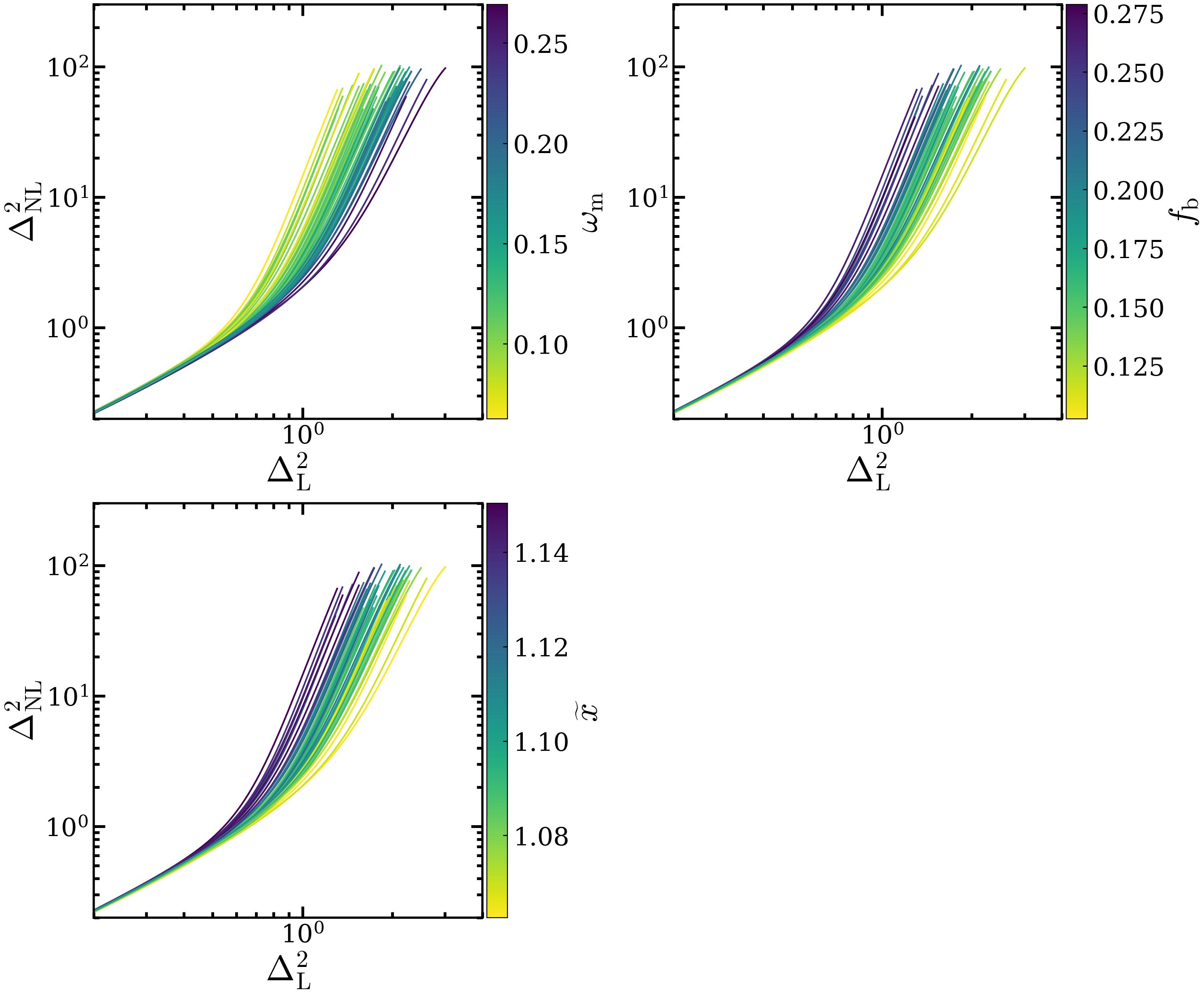}
    \caption{Dimensionless power spectrum coloured according to their values of: physical matter density $\omega_\mathrm{m}$ and baryon fraction $f_\mathrm{b}$ (Row 1) and $\tilde{x}$ (Row 2 ) respectively. All of the plots are for cosmologies within 20$\sigma$ of Planck as produced by \spinex.}
    \label{fig:how_params_change_SPINEX_mainpaper}
\end{figure}

Similarly, Fig.\ref{fig:how_params_change_SPINEX_mainpaper} presents the dimensionless power spectra of \spinex for within $20\sigma$ of the Planck measurements. This model exhibits a comparable trend in relation to $\omega_\mathrm{m}$ and $f_\mathrm{b}$, with the observed behaviour in $f_b = \omega_\mathrm{b}/\omega_\mathrm{m}$ being directly influenced by $\Omega_\mathrm{m}$. Large values of $\tilde{x}$ make $\Delta^2(k)$ go nonlinear earlier, as can be seen in Fig.\ref{fig:how_params_change_SPINEX_mainpaper}. This is the opposite of what we see for large values of $g_a$. As $\tilde{x}$ is the integral of the growth history in time intervals of $\tau = \mathrm{ln} (\sigma_{12})$, we can see that these parameters are related as $\tilde{x} \propto 1/g_a$. This phenomenon can also be explained in the context of growth dynamics: $g_a$ quantifies growth suppression relative to an Einstein-de Sitter universe, while $\tilde{x}$ encapsulates insights regarding structure growth histories. Consequently, increased growth suppression correlates with diminished structural growth, establishing an inverse relationship between these two variables.

When we compare the MAPE of the two models, we see that the \spinex parameterisation performs better than that of \spine for two of the scenarios. As previously mentioned in section \ref{sec:methodology_power_spectrum}, the variables used for \spinex capture the shape of the power spectrum with greater accuracy, and they have the advantage of being independent of $h$ \citep[see][for a review]{arguments_h, evolution_mapping}. 

\subsection{Model Performance} \label{sec:discussion_model_performance}
The performance of our model is consistent with that of other power spectrum predictors. However, we must note the `wiggles' around $k\sim0.1\;h\;\mathrm{Mpc}^{-1}$ in the residual plots for all power spectrum fits. The most likely culprit for this is the initialisation of Fourier modes in the Quijote simulation suite. The set of simulations used in this work have fixed initial conditions with the same initial random seed. Although the reduction in cosmic variance through fixed and paired-fixed simulations is excellent, it is limited to only the long-wavelength modes ($k\lesssim 0.05 \; h\;\mathrm{Mpc}^{-1}$) \citep{suppressing_cosmic_variance}. The effects of pairing and fixing are weak even in the weakly nonlinear regime ($k\gtrsim 0.1 \; h\;\mathrm{Mpc}^{-1}$). These two effects can be seen in all the fits for $P_{\mathrm{NL}}(k_{\mathrm{NL}})$ throughout this work. It is therefore safe to assume that the decrease in accuracy around $k\sim 0.1 \; h\;\mathrm{Mpc}^{-1}$ for both \spine and \spinex comes from this phenomenon. To further confirm this, we have compared the fixed-seed Quijote LH simulations to other emulators in Appendix \ref{sec:appendixB} to demonstrate that wiggles with a similar shape appear at the same scales.

Another source of uncertainty arises from the dewiggling process, as this method demonstrates varying effectiveness depending on the specific cosmological framework. In particular, certain cosmologies exhibit significantly large baryon oscillations, which result in residual wiggles when applying the dewiggling prescription. PD96 do state that the assumption presented in equation~\eqref{eqn:pd_mapping_2} fails for models with oscillations typically associated with high baryons. This phenomenon could be the cause of the diminished accuracy of models that incorporate high baryon conditions. Despite this, our model continues to be consistent with other emulators and offers a satisfactory fit for the power spectrum. 

Remarkably, even with only 130 cosmologies, PySR has been able to derive symbolic expressions for the power spectrum that demonstrate sub-5\% accuracy up to $k=1\;h\;\mathrm{Mpc}^{-1}$. When compared to some of the emulators used in this study, our emulators possess the advantage of a broader parameter range, which makes them useful for exploring more exotic models. For the LH cosmologies, SYREN-Halofit likewise extends beyond its defined parameter range. Given the performance of SYREN-Halofit in this domain, it can be concluded that symbolic emulators exhibit considerable potential for extrapolating beyond their training range while also maintaining high levels of accuracy.

\begin{figure}
    \centering
    \includegraphics[width=\columnwidth]{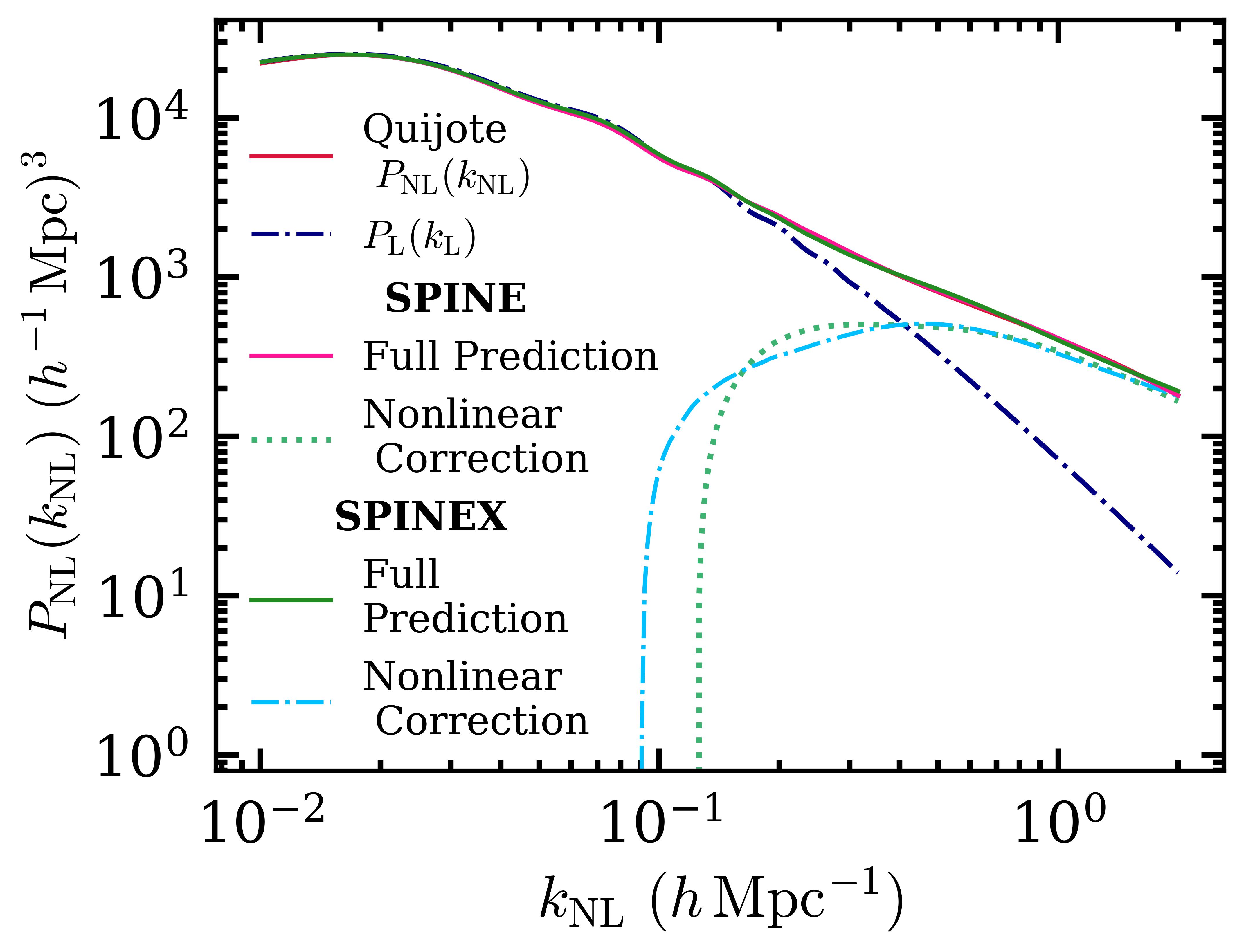}
    \caption{The effect of the two terms in both of our symbolic equations as plotted for the Quijote fiducial parameters (see Table \ref{tab:Quijote_limits}). The power spectra shown here are the full predictions, i.e., with the BAO signal fully reintegrated. This formulation works as intended. The first term in both models, depicting the linear power spectrum (singular navy dash-dotted line for both \spine and \spinex as they are identical). The nonlinear correction term, i.e., equations~\eqref{eqn:spine_final_eqn} and \eqref{eqn:spinex_final_eqn}, for \spine (green dotted line) and \spinex (blue dash-dotted line) is non-existent at large scales and only starts to come into effect at intermediate scales, eventually dominating the equations deep in the nonlinear regime. We have also plotted the Quijote fiducial nonlinear power spectrum $P_{\mathrm{NL}}(k_{\mathrm{NL}})$ (red solid line) for comparison.}
    \label{fig:effect_of_two_terms}
\end{figure}

We also highlight that our choice to decompose the nonlinear power spectrum into a linear contribution with an additional nonlinear correction also operates as intended. In Fig.\ref{fig:effect_of_two_terms}, we plot these two terms separately for the \spine and \spinex predictions for the nonlinear power spectrum. We can clearly see that the linear term dominates at large-scales with the nonlinear correction term - $N\left[\Delta_{\mathrm{L}}^2(k_{\mathrm{L}}),\theta \; \mathrm{or} \; \theta_\mathrm{X}\right]$ vanishing at these scales as expected. As nonlinearities start to become significant, $N\left[\Delta_{\mathrm{L}}^2(k_{\mathrm{L}}),\theta \; \mathrm{or} \; \theta_\mathrm{X}\right]$ begin to take precedence, ultimately becoming dominant in the deep nonlinear regime. Since the nonlinear correction terms vanish for small $k$, the extrapolation at large scales is the linear power spectrum itself. We do not recommend extrapolating to $k$ values beyond the scales examined in this work.

\subsection{Evolution of Equations}
In this section, we examine the progression of the functional form of the equations spawned at each complexity for \spine and \spinex. This is to aid in understanding the origins of the final equations from SR. 

We present some key equations generated for the \spine model in Table \ref{tab:spine_eqn_table}. Upon examining equations 1 through 7, one can identify a wide array of functional forms. These forms do not serve as viable candidates as they do not fit the data well. Their corresponding fits are shown in Fig.\ref{fig:all_eqn_plot}. Interestingly, equations 3 and 4 incorporate $\Omega_\mathrm{m}$ and $g_\mathrm{a}$, inter-connected quantities that were previously shown to impact the growth of structure.

In equation number 7, the combination $c^{n_{\mathrm{L}}}g_\mathrm{a}$ (where $c$ denotes a numerical constant) is established. This term is retained in the final fit, as represented in equation \ref{eqn:spine_final_eqn}. Furthermore, terms that incorporate $\Delta^2_{\mathrm{L}}(k_{\mathrm{L}})$ with decimal exponents are crucial, as their inclusion enhances the overall fit of the function. This term first appears in equation 2. The larger the value of $\Delta^2_{\mathrm{L}}(k_{\mathrm{L}})$, the larger the nonlinear correction  $N\left[\Delta^2_{\mathrm{L}}(k_{\mathrm{L}}),\theta\right]$ becomes; this is as expected as larger $\Delta^2_{\mathrm{L}}(k_{\mathrm{L}})$ corresponds to more nonlinear scales. This exponent is calibrated at each complexity level to improve the fit.

Equations 8 through 14 provide adjustments to equation 7, enhancing its accuracy in incremental ways. One can observe these refinements through the gradual decline in loss on the Pareto front around complexity 13 (see Fig.\ref{fig:pareto_curve}). In equation 15, the incorporation of $\sigma_8$ and the inclusion of $\Delta^2_{\mathrm{L}}(k_{\mathrm{L}})$ as an exponent yield a notable improvement in accuracy. Although this approach may appear less intuitive, terms with this function appear to improve the precision of fit. Furthermore, equation 16 introduces an additional multiplicative factor of $\Delta^2_{\mathrm{L}}(k_{\mathrm{L}})$, a characteristic that is preserved in the final function and contributes to its overall accuracy. The dip in the Pareto curve presents this rise in accuracy around complexity 31. 

Equations 16 through 26 integrate adjustments that gradually improve accuracy. These adjustments involve the incorporation of various terms that consist of combinations of $\Omega_\mathrm{m}$ and $n_\mathrm{s}$. Furthermore, constants are refined to achieve a more precise fit. Equation 24 is interesting, where the denominator $\sigma_8 g_\mathrm{a}$ is introduced. Despite its presence in the equation, this inclusion does not yield a significant improvement in accuracy, as we would expect. This observation implies that the term primarily improves accuracy moderately rather than significantly. In summary, these equations exhibit strong similarities to the final functional form outlined in equation \ref{eqn:spine_final_eqn}.

Equations 27 through 39 represent the Pareto plateau, with equation 27 designated as the final selection for fitting the dimensionless power spectrum (given by equation \ref{eqn:spine_final_eqn} in this work). The introduction of the term $-3.76/\Delta^2_{\mathrm{L}}(k_{\mathrm{L}})$ is substantial, as it significantly improves the fit and is responsible for the observed dip in loss at the onset of the Pareto plateau (see Fig.\ref{fig:pareto_curve}), around complexity 53. Subsequent modifications to the functional form, starting from equation 27, tend to complicate the equation without providing a substantial improvement in fit.

To summarise, the incorporation of terms of the form $c^{n_{\mathrm{L}}} g_{\mathrm{a}}$, utilizing $\Delta^2_{\mathrm{L}}(k_{\mathrm{L}})$ as an exponent, along with the addition of $\sigma_8$ to parametrise the power spectrum, results in a significant improvement in the fit. The calibration of constants and the integration of various terms, including $\sigma_8 g_{\mathrm{a}}$, represent modifications aimed at refining the overall accuracy of the power spectrum.

The same approach can be applied to equations obtained from \spinex to get a better understanding of how its functional form has originated. Some key equations obtained for \spinex can be found in Table \ref{tab:spinex_eqn_table}. Immediately, we see that the first few equations generated for both models have identical functional forms. Furthermore, we see the inclusion of $\Delta^2_{\mathrm{L}}(k_{\mathrm{L}})$ with decimal exponents as well. 

Equation 7 onwards, the functional form is further refined. The loss dramatically drops for equation 15 as can be seen by \spinex's Pareto curve in Fig.\ref{fig:pareto_curve}. The spectral index, $n_\mathrm{s}$, appearing in the denominator is a favoured combination and is retained in the final chosen functional form as well, as given by equation \ref{eqn:spinex_final_eqn} in this work. At each iteration, constants are recalibrated, and the improvement in loss becomes more gradual following Equation 16. The final equation that makes up \spinex is equation 27, provided in equation \ref{eqn:spinex_final_eqn} in this work. This equation was chosen by eye as it resides approximately at the beginning of the Pareto plateau for this model. After this equation, different terms and constants are added, which increase the complexity of the model and do not contribute much to the improvement in loss. 

Although the lower complexity equations for both models offer simpler solutions, they do not reproduce the power spectrum with the same level of accuracy as the ones found further down the Pareto front. Lower-complexity models consistently fail to capture the shape of the power spectrum and give rise to larger errors. As previously discussed in Section \ref{sec:results_auxilliary_eqns}, this behaviour can also be seen in Fig.\ref{fig:all_eqn_plot} top panel for the equations generated for \spine. To provide a good fit for the power spectrum, we therefore opted to choose more complex equations despite some terms in the equations not being readily interpretable and not having any clear physical motivation.

\begin{table*}
\caption{Table of a few key equations obtained for the \spine model. The columns provide equation number, equation complexity and the nonlinear correction term for that corresponding complexity as a function of $\Delta^2_{\mathrm{L}}(k_{\mathrm{L}})$ and $\theta$. See Table \ref{tab:model_params} for more details.}
    \centering
    \begin{tabular}{p{0.01\textwidth} p{0.06\textwidth} p{0.82\textwidth}}
    \hline
    \\
    \textbf{Eq.} & \textbf{Complexity} & \textbf{Nonlinear Correction Term $N\left[\Delta_{\mathrm{L}}^2(k_{\mathrm{L}}), \theta \right]$} \\
    \\
    \hline
    \\
        1 & 1 & $14.155$ \\
        2 & 3 & ${5.988}^{\Delta_\mathrm{L}^2(k_\mathrm{L})}$ \\
        3 & 5 & $\frac{\left(\Delta_\mathrm{L}^2(k_\mathrm{L})\right)^{3.721}}{\Omega_\mathrm{m}}$ \\
        4 & 7 & $\left(\Delta_\mathrm{L}^2(k_\mathrm{L}) g_\mathrm{a}^{n_\mathrm{L}}\right)^{3.459}$ \\
        5 & 9 & $0.018\cdot {0.042}^{n_\mathrm{L}} \left(\Delta_\mathrm{L}^2(k_\mathrm{L})\right)^{3.545}$ \\
        6 & 11 & $\left(\Delta_\mathrm{L}^2(k_\mathrm{L})\right)^{3.671} \left(5.363 \cdot{4.099}^{n_\mathrm{L}}\right)^{n_\mathrm{L}}$ \\
        7 & 13 & $\left(\Delta_\mathrm{L}^2(k_\mathrm{L})\right)^{3.871} \left(5.679\cdot {3.576}^{n_\mathrm{L}} g_\mathrm{a}\right)^{n_\mathrm{L}}$ \\
        8 & 15 & $\left(\Delta_\mathrm{L}^2(k_\mathrm{L})\right)^{3.919} \left(4.681\cdot {2.911}^{n_\mathrm{L}} g_\mathrm{a} - 0.075\right)^{n_\mathrm{L}}$ \\
        $\cdot$ & $\cdot$ & $\cdot$ \\
        $\cdot$ & $\cdot$ & $\cdot$ \\
        $\cdot$ & $\cdot$ & $\cdot$ \\
        14 & 27 & $\left(\Delta_\mathrm{L}^2(k_\mathrm{L})\right)^{4.288} \left(\left({3.89}^{n_\mathrm{L}} \left(g_\mathrm{a} \left(h + 9.362\right) + n_\mathrm{L}\right)\right)^{n_\mathrm{L}} + \left(\Delta_\mathrm{L}^2(k_\mathrm{L}) \left(n_\mathrm{s}^{\Delta_\mathrm{L}^2(k_\mathrm{L})} + 0.251\right)\right)^{-2.324}\right)$ \\
        15 & 29 & $\left(\Delta_\mathrm{L}^2(k_\mathrm{L})\right)^{3.045} \left(- n_\mathrm{L} + \left({4.311}^{n_\mathrm{L}} g_\mathrm{a} \left(\left(- n_\mathrm{L} + 745.376 \left(\frac{{15.359}^{n_\mathrm{L}}}{\sigma_{8}}\right)^{\Delta_\mathrm{L}^2(k_\mathrm{L})}\right)^{\Delta_\mathrm{L}^2(k_\mathrm{L})} + 3.961\right)\right)^{n_\mathrm{L}}\right)$ \\
        16 & 31 & $\left(\Delta_\mathrm{L}^2(k_\mathrm{L})\right)^{2.638} \left(\Delta_\mathrm{L}^2(k_\mathrm{L}) \left({4.952}^{n_\mathrm{L}} g_\mathrm{a} \left(\left(- n_\mathrm{L} + 638.035 \left(\frac{{14.777}^{n_\mathrm{L}}}{\sigma_{8}}\right)^{\Delta_\mathrm{L}^2(k_\mathrm{L})}\right)^{\Delta_\mathrm{L}^2(k_\mathrm{L})} + 7.4251394\right)\right)^{n_\mathrm{L}} - n_\mathrm{L}\right)$ \\
        $\cdot$ & $\cdot$ & $\cdot$ \\
        $\cdot$ & $\cdot$ & $\cdot$ \\
        $\cdot$ & $\cdot$ & $\cdot$ \\
        24 & 47 & \begin{multline} \Bigl(\Delta_\mathrm{L}^2(k_\mathrm{L})\Bigr)^{2.726} \Bigl(\Delta_\mathrm{L}^2(k_\mathrm{L}) \Bigl(\Bigl({4.021}^{n_\mathrm{L}} g_\mathrm{a}^{n_\mathrm{s}} \left(\Bigl(- n_\mathrm{L} + \frac{\left({9.416}^{n_\mathrm{L}} n_\mathrm{s}\right)^{\Delta_\mathrm{L}^2(k_\mathrm{L})} \left(169.048 - \frac{4.648}{\Omega_\mathrm{m} - \Omega_\mathrm{b}}\right)}{\sigma_{8} g_\mathrm{a}}\right)^{\Delta_\mathrm{L}^2(k_\mathrm{L})} \\ + 4.809\Bigr) - 0.046\Bigr)^{n_\mathrm{L}} - 0.173\Bigr) - n_\mathrm{L}\Bigr) \nonumber \end{multline} \\
        $\cdot$ & $\cdot$ & $\cdot$ \\
        $\cdot$ & $\cdot$ & $\cdot$ \\
        $\cdot$ & $\cdot$ & $\cdot$ \\
        27 & 53 & $\left(\Delta_\mathrm{L}^2(k_\mathrm{L})\right)^{2.644} \left(\Delta_\mathrm{L}^2(k_\mathrm{L}) \left(\left({4.164}^{n_\mathrm{L}} g_\mathrm{a}^{n_\mathrm{s}} \left(\left(- n_\mathrm{L} + \frac{\left({9.13}^{n_\mathrm{L}} n_\mathrm{s}\right)^{\Delta_\mathrm{L}^2(k_\mathrm{L})} \left(162.451 + \frac{n_\mathrm{L}^{3} + \frac{3.76}{\Delta_\mathrm{L}^2(k_\mathrm{L})}}{\Omega_\mathrm{m} - \Omega_\mathrm{b}}\right)}{\sigma_{8} g_\mathrm{a}}\right)^{\Delta_\mathrm{L}^2(k_\mathrm{L})} + 
        4.241\right)\right)^{n_\mathrm{L}} - 0.162\right) - n_\mathrm{L}\right)$ \\
        $\cdot$ & $\cdot$ & $\cdot$ \\
        $\cdot$ & $\cdot$ & $\cdot$ \\
        $\cdot$ & $\cdot$ & $\cdot$ \\
        \\
        \hline
    \end{tabular}
    \label{tab:spine_eqn_table}
\end{table*}

\begin{table*}
\caption{Table for some key equations obtained for the \spinex models. The columns provide equation number, equation complexity and the nonlinear correction term for that corresponding complexity as a function of $\Delta^2_{\mathrm{L}}(k_{\mathrm{L}})$ and $\theta_X$. See Table \ref{tab:model_params} for more details.}
    \centering
    \begin{tabular}{p{0.01\textwidth} p{0.06\textwidth} p{0.82\textwidth}}
    \hline
    \\
    \textbf{Eq.} & \textbf{Complexity} & \textbf{Nonlinear Correction Term $N\left[\Delta_{\mathrm{L}}^2(k_{\mathrm{L}}), \theta_X \right]$} \\
    \\
    \hline
    \\
        1 & 1 & $15.089$ \\
        2 & 3 & ${6.010}^{\Delta^2_{\mathrm{L}}(k_{\mathrm{L}})}$ \\
        3 & 5 & $\frac{\left(\Delta^2_{\mathrm{L}}(k_{\mathrm{L}})\right)^{2.942}}{\omega_\mathrm{m}}$ \\
        4 & 7 & $f_\mathrm{b} \left({0.168}^{n_\mathrm{L}}\right)^{\Delta^2_{\mathrm{L}}(k_{\mathrm{L}})}$ \\
        5 & 9 & $0.014\cdot {0.037}^{n_\mathrm{L}} \left(\Delta^2_{\mathrm{L}}(k_{\mathrm{L}})\right)^{3.549}$ \\
        6 & 11 & $\left(\Delta^2_{\mathrm{L}}(k_{\mathrm{L}})\right)^{3.672} \left(5.917\cdot {4.345}^{n_\mathrm{L}}\right)^{n_\mathrm{L}}$ \\
        7 & 13 & $\left(\Delta^2_{\mathrm{L}}(k_{\mathrm{L}})\right)^{3.875} \left(4.908 \left({3.342}^{\tilde{x}}\right)^{n_\mathrm{L}}\right)^{n_\mathrm{L}}$ \\
        $\cdot$ & $\cdot$ & $\cdot$ \\
        $\cdot$ & $\cdot$ & $\cdot$ \\
        $\cdot$ & $\cdot$ & $\cdot$ \\
        15 & 29 & $\Delta^2_{\mathrm{L}}(k_{\mathrm{L}}) \left(\Delta^2_{\mathrm{L}}(k_{\mathrm{L}}) + \frac{0.075 \Delta^2_{\mathrm{L}}(k_{\mathrm{L}})^{4.227} \left(\left(\left(\left(- 0.614 \Delta^2_{\mathrm{L}}(k_{\mathrm{L}}) + n_\mathrm{s} + \tilde{x}\right)^{\sigma_{12}} + 1.432\right)^{\tilde{x}}\right)^{n_\mathrm{L}}\right)^{n_\mathrm{L}}}{n_\mathrm{s}}\right)$ \\
        $\cdot$ & $\cdot$ & $\cdot$ \\
        $\cdot$ & $\cdot$ & $\cdot$ \\
        $\cdot$ & $\cdot$ & $\cdot$ \\
        27 & 53 & $\frac{\left(\Delta^2_{\mathrm{L}}(k_{\mathrm{L}})\right)^{4.305} \left(0.085 \Delta^2_{\mathrm{L}}(k_{\mathrm{L}}) \left(\left(\left(\tilde{x} \left(f_\mathrm{b} + \tilde{x} + \left(- 0.652 \Delta^2_{\mathrm{L}}(k_{\mathrm{L}}) \tilde{x} + f_\mathrm{b}^{-0.153} + n_\mathrm{s}^{\tilde{x}}\right)^{\sigma_{12}}\right)\right)^{n_\mathrm{L}}\right)^{n_\mathrm{L}} + \frac{f_\mathrm{b} - \tilde{x}^{-15.169}}{\sigma_{12}}\right) + \frac{f_\mathrm{b} + \tilde{x} - 0.339}{\Delta^2_{\mathrm{L}}(k_{\mathrm{L}})^{2}}\right)}{n_\mathrm{s}}$\\
        $\cdot$ & $\cdot$ & $\cdot$ \\
        $\cdot$ & $\cdot$ & $\cdot$ \\
        $\cdot$ & $\cdot$ & $\cdot$ \\
        \\
        \hline
    \end{tabular}
    \label{tab:spinex_eqn_table}
\end{table*}

\section{Conclusion} \label{sec:conclusion}
In this paper, we presented \spine and \spinex, two symbolic models developed for the calculation of the $z=0$ nonlinear matter power spectrum $P_{\mathrm{NL}}(k_{\mathrm{NL}})$ within the $\Lambda$CDM framework, assuming $M_{\nu}=0$. This approach aims to mitigate the limitations associated with N-body simulations and other numerical emulation techniques regarding computational efficiency and speed. Furthermore, it seeks to enhance the understanding of nonlinear evolution and the parameters most important in that context. The foundation of this research is based on the mapping established in \cite{PD96} and seeks to express $P_{\mathrm{NL}}(k_{\mathrm{NL}})$ in relation to its linear counterpart and several cosmological parameters through well-defined mathematical equations within the interval $0.01\;h\;\mathrm{Mpc}^{-1}<k<2 \;h\;\mathrm{Mpc}^{-1}$. Both models have a parameter range as outlined in Table \ref{tab:Quijote_limits}. These equations are derived through symbolic regression via genetic programming as implemented by PySR \citep{pysr}.

The principal distinctions between the two models discussed in this paper are fundamentally related to the parameters they encompass. The model \spine (equation~\ref{eqn:spine_final_eqn}) seeks to characterise $P_{\mathrm{NL}}(k_{\mathrm{NL}})$ with parameters $\theta = \left \{\Omega_\mathrm{m}, \Omega_\mathrm{b}, h, n_\mathrm{s}, \sigma_8, n_\mathrm{L}, g_\mathrm{a} \right \}$. In contrast, \spinex (equation~\ref{eqn:spinex_final_eqn}) utilises the parameters $\theta_\mathrm{X}= \left\{\omega_\mathrm{m}, f_\mathrm{b}, n_s, \sigma_{12}, n_\mathrm{L}, \tilde{x} \right\}$. The $\theta_X$ parameterisation is characterised as using physical parameters, i.e. parameters that avoid any explicit dependency on $h$.

Removal of the BAO signal enhances the accuracy with which we can capture the overall broadband shape of the power spectrum; hence, the equations that make up \spine and \spinex are both designed to produce a `dewiggled' version of the dimensionless power spectrum $\Delta^2_{\mathrm{NL}}(k_{\mathrm{NL}})$ (see section \ref{sec:methodology_bao}). However, one can easily obtain $P_{\mathrm{NL}}(k_{\mathrm{NL}})$ by following the pipeline presented in section \ref{sec:methodology_calculate_pk}. Furthermore, we constructed equations that incorporate a linear power spectrum and a nonlinear correction term to enhance the understanding of the parameters that most significantly guide nonlinear evolution.

We validate both models against our test set of cosmologies, obtained from the Quijote LH and fiducial simulations \citep{quijote_simulations} for three different cases. Firstly, we obtain a prediction for cosmologies that are within 20$\sigma$ of the \citealt{planck_2018} observations. In this case, \spine has a mean absolute percentage error (MAPE) of \PLANCKmapeSPINE\% while \spinex is \PLANCKmapeSPINEX\%. Secondly, for a designated fiducial cosmology, \spine has a MAPE of \FIDmapeSPINE\% and \spinex is \FIDmapeSPINEX\%. Finally, we also wanted to obtain a prediction for a more general case. Here, we use randomly selected LH cosmologies. These cosmologies embody a large parameter range. \spine produces a MAPE of \LHmapeSPINE\% while \spinex is \LHmapeSPINEX\%. A summary of the MAPE for all models is given in Table \ref{tab:mape_models}. Overall, we obtain a good fit for all $k$-scales analysed here. We also compare our models to other emulators and find that our models are consistent with those analysed here. The benefit of having symbolic expressions to describe a quantity is their superior extrapolation behaviour. Due to this, the one advantage we have is that we can produce $P_{\mathrm{NL}}(k_{\mathrm{NL}})$ for a broad range of cosmologies with sub-5\% accuracy in a majority of cases. 

In the absence of constraints related to computational power and memory limitations, a more precise equation could have been developed using the complete set of cosmologies available in the Quijote suite, perhaps even utilising higher-resolution simulations. However, this would have introduced additional complexity and computational burden. Variables and operators for specific scales could be introduced; however, in this work, we have aimed for simplicity and explainability, and this approach to fitting the power spectrum through symbolic equations serves as a proof of concept and holds the potential to facilitate the development of analytical models that provide predictions for $P_{\mathrm{NL}}(k_{\mathrm{NL}})$ across multiple redshifts, as well as redshift-space power spectra. This would also involve the construction of a comprehensive framework for bias and redshift space distortions, thereby preparing for the subsequent application of these models to actual survey data. In future work, it will be of significant interest to examine how the equations may be modified with the introduction of smaller scales. One could also use the output from other emulators, such as EuclidEmulator, to develop a symbolic model. Furthermore, although this work primarily examines the $\Lambda$CDM model, future work can utilise the \spinex model for predicting $P_{\mathrm{NL}}(k_{\mathrm{NL}})$ in cosmologies characterised by dynamical dark energy, to see how well $\tilde{x}$ captures the influence of dynamical dark energy on the growth of structure.

\section{Acknowledgements} \label{sec:acknowledgements}
The authors of this paper would like to extend their gratitude to Dr Miles Cranmer for their invaluable input and assistance through the PySR GitHub platform. We would also like to thank Dr Deaglan J. Bartlett for their useful suggestions. We also acknowledge the Viper High Performance Computing facility of the University of Hull and its support team.

\section{Data Availability}
We have created a Python implementation of both \spine and \spinex. This code is publicly available through the \texttt{spine-pk}\footnote{\url{https://github.com/skadoosh-MC/SPINE.git}} python package.

\FloatBarrier
\bibliographystyle{mnras}
\bibliography{references}

\appendix
\section{Additional Parameters} \label{appendixA:rest_of_the_params}
\begin{figure}
    \centering
    \includegraphics[width=\columnwidth]{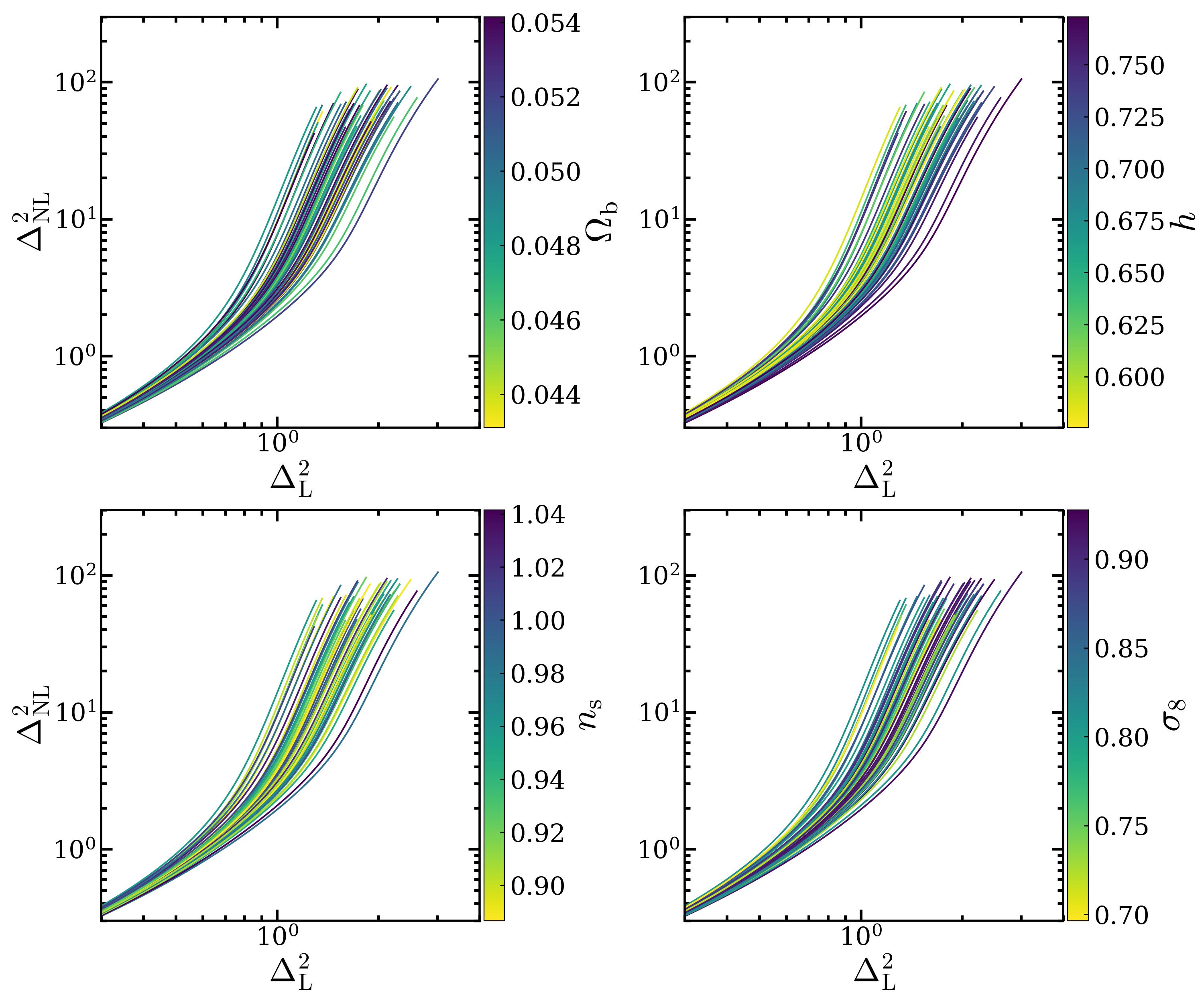}
    \caption{Dimensionless power spectra coloured according to their values of: baryon density $\Omega_\mathrm{b}$ and $h$ (Row 1) and $n_\mathrm{s}$ and $\sigma_8$ (Row 2) respectively. All of the plots are for cosmologies that are within 20$\sigma$ of Planck-2018 as produced by \spine.}
    \label{fig:how_params_change_spine_appendix}
\end{figure}

\begin{figure}
    \centering
    \includegraphics[width=\columnwidth]{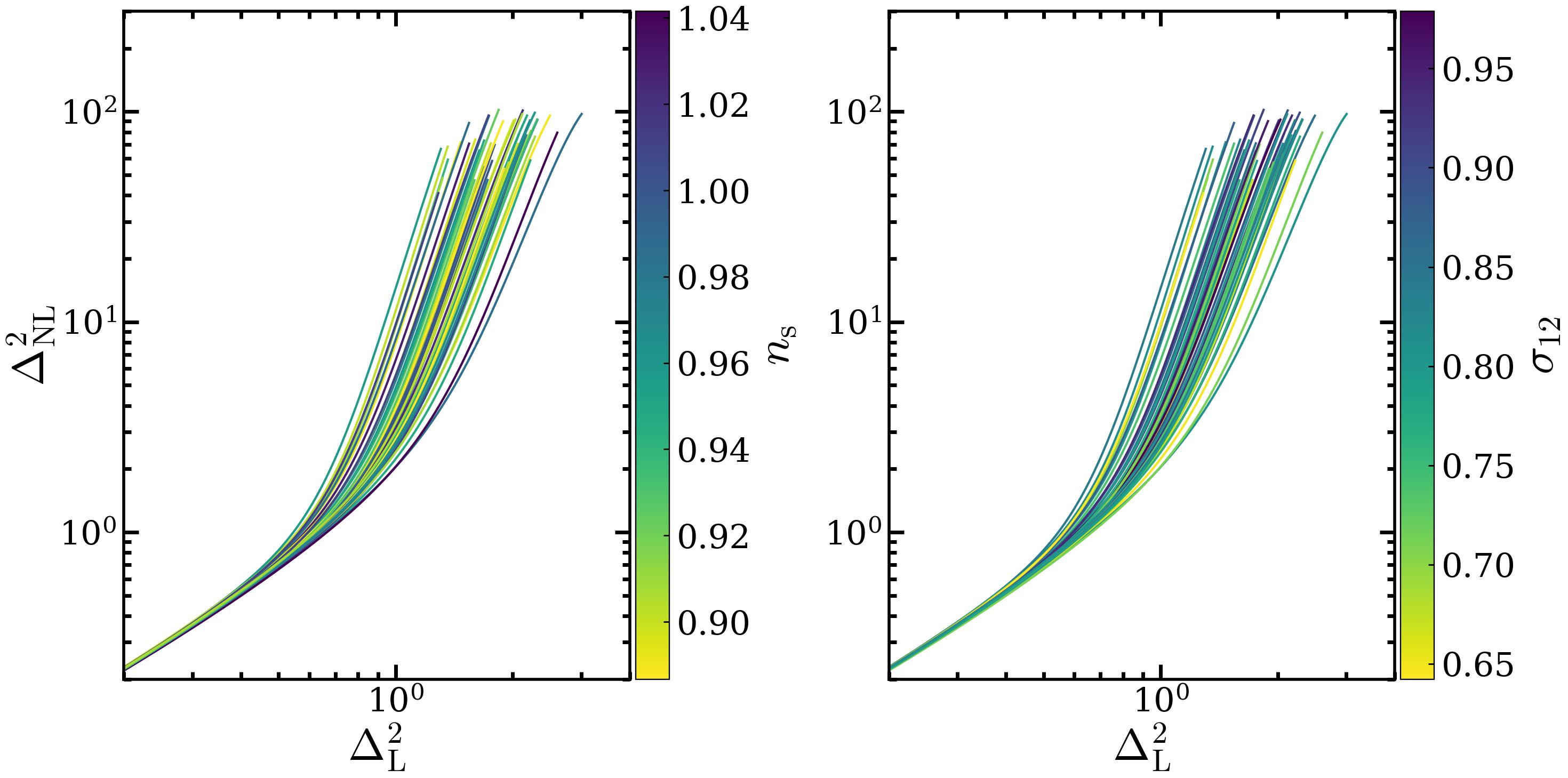}
    \caption{Dimensionless power spectrum plots coloured according to their values of $n_\mathrm{s}$ and $\sigma_{12}$ respectively. All plots are for cosmologies that are within 20$\sigma$ of the Planck-2018 observations as produced by \spinex.}
    \label{fig:how_params_change_SPINEX_appendix}
\end{figure}

Figs. \ref{fig:how_params_change_spine_appendix} and \ref{fig:how_params_change_SPINEX_appendix} depict the parameters, which visually have no obvious pattern when it comes to $\Delta^2(k)$.

\section{Cosmic Variance at Intermediate Scales}\label{sec:appendixB}

\begin{figure*}
\centering
\begin{subfigure}[t]{0.48\textwidth}
    \centering
    \includegraphics[width=\linewidth]{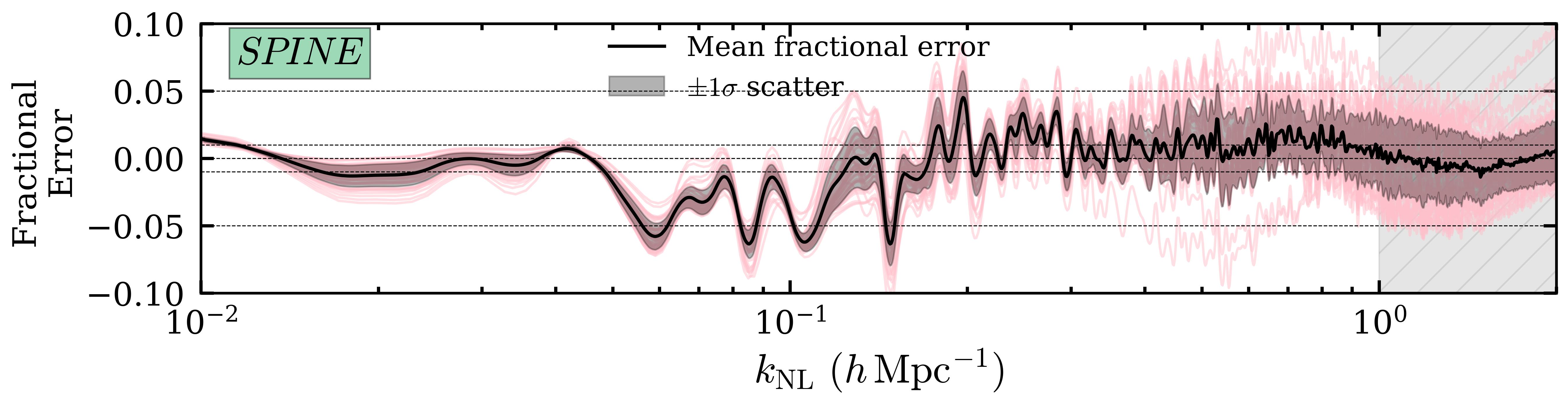}
    \caption{}
    \label{subfig:planck_spine_residuals_appendix}

    \vspace{0.3cm}

    \includegraphics[width=\linewidth]{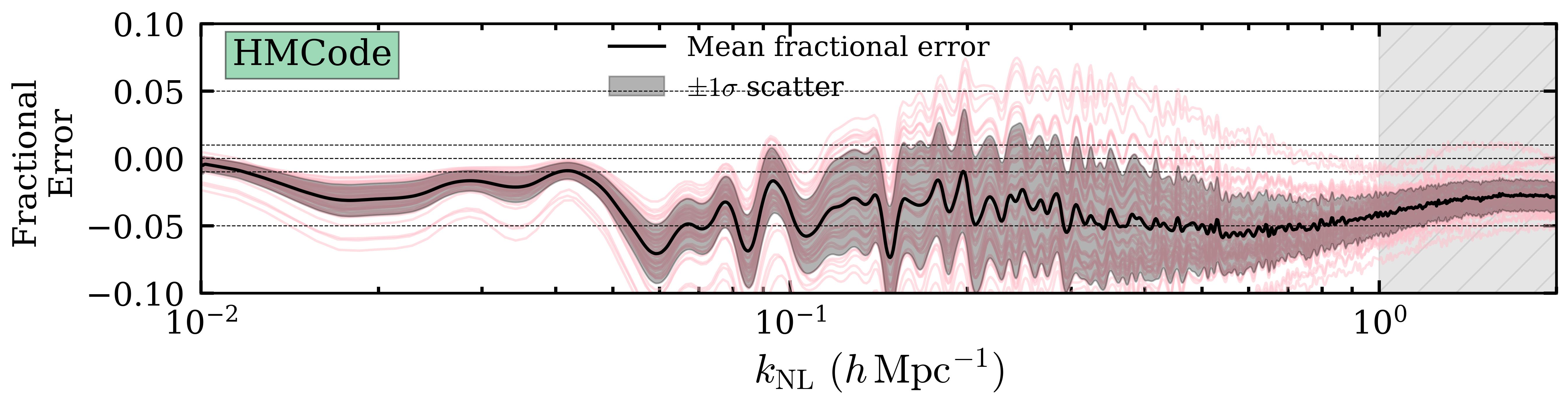}
    \caption{}
    \label{subfig:planck_hmcode_residuals}

    \vspace{0.3cm}

    \includegraphics[width=\linewidth]{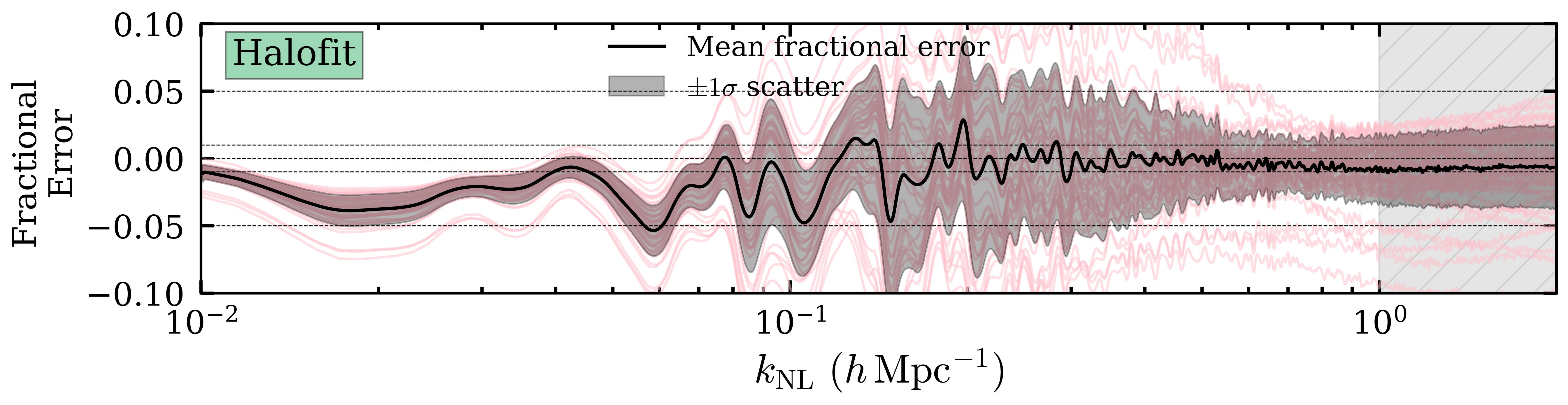}
    \caption{}
    \label{subfig:planck_halofit_residuals}
\end{subfigure}
\hfill
\begin{subfigure}[t]{0.48\textwidth}
    \centering
    \includegraphics[width=\linewidth]{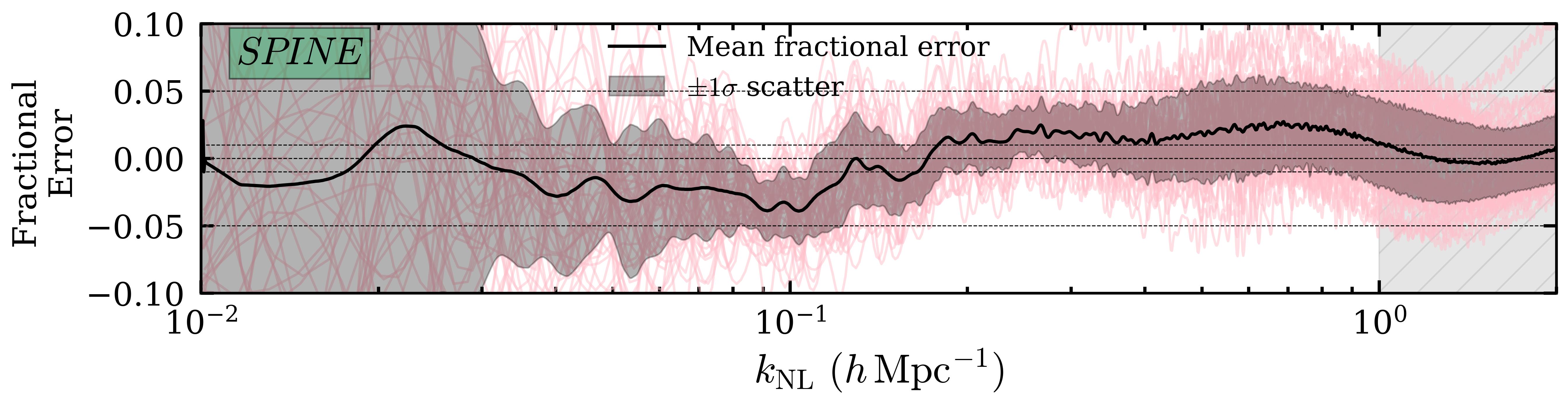}
    \caption{}
    \label{subfig:planck_spine_standard_residuals_appendix}

    \vspace{0.3cm}

    \includegraphics[width=\linewidth]{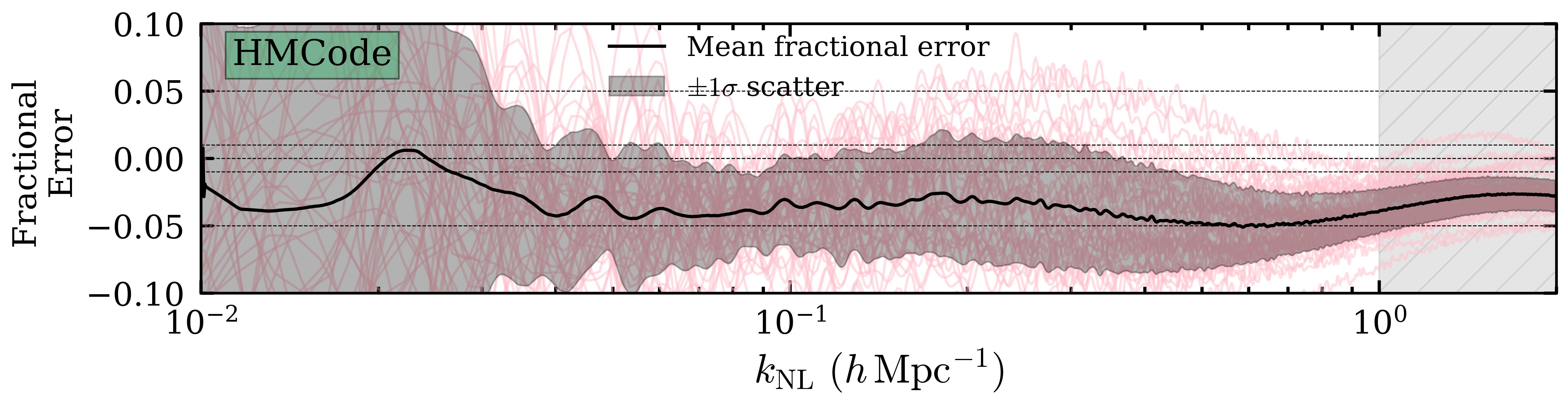}
    \caption{}
    \label{subfig:planck_standard_hmcode_residuals}

    \vspace{0.3cm}

    \includegraphics[width=\linewidth]{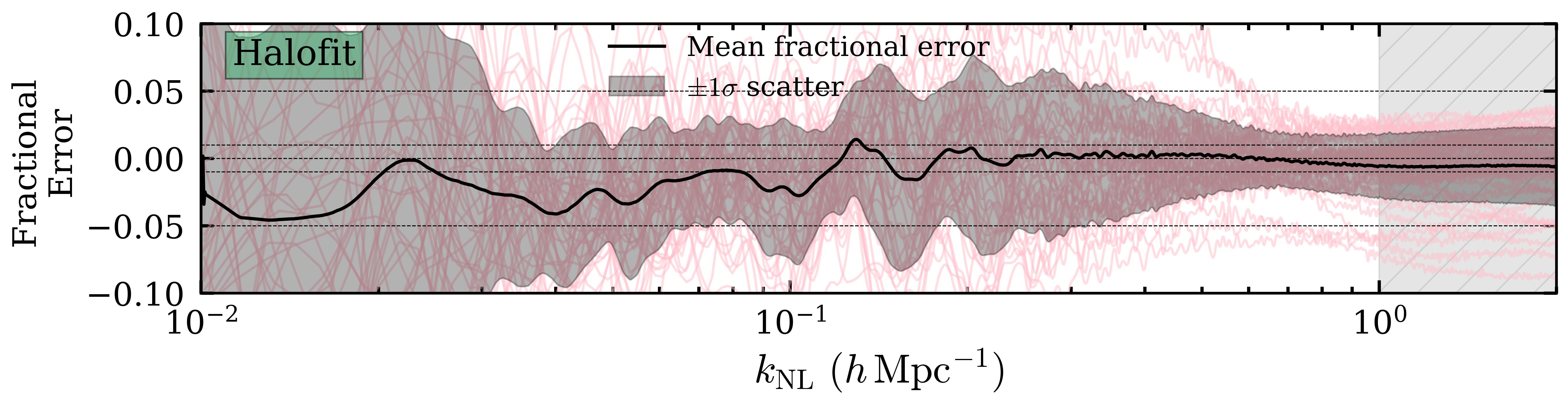}
    \caption{}
    \label{subfig:planck_standard_halofit_residuals}
\end{subfigure}

\caption{\textbf{(a--c):} Residuals for the paired-fixed Planck-like cosmologies using \spine\ (a), HMCode (b), and Halofit (Takahashi) (c). \textbf{(d--f):} Corresponding residuals for the standard Planck-like cosmologies using \spine\ (d), HMCode (e), and Halofit (Takahashi) (f). All residuals are computed with respect to the Quijote nonlinear power spectrum. The black line in each panel shows the mean fractional error, while the shaded region denotes the $\pm1\sigma$ scatter. The residual oscillations around $k\sim0.1\,h\,\mathrm{Mpc}^{-1}$ are signatures of residual cosmic variance and are visible for all three emulators in the case of fixed simulations.}
\label{fig:spine_halofit_hmcode}
\end{figure*}

In cosmological simulations, cosmic variance manifests as realisation-to-realisation scatter resulting from different random initial conditions. This scatter becomes increasingly important on large spatial scales, where the number of independent Fourier modes is limited. Fixed and paired fixed simulations mitigate this (see Section \ref{sec:data}). However, as we have stated in Section \ref{sec:discussion_model_performance}, residual cosmic variance still exists at scales around $k\sim0.1\;h\;\mathrm{Mpc}^{-1}$. It is important to test that the fluctuations observed at this scale are indeed attributed to cosmic variance and not to any fitting discrepancies associated with either \spine or \spinex.

As verification, we illustrate this phenomenon by comparing the residuals of \spine (Fig. \ref{subfig:planck_spine_residuals_appendix}), HMCode (Fig. \ref{subfig:planck_hmcode_residuals}), and Halofit (Fig. \ref{subfig:planck_halofit_residuals}). For comparison, we have also plotted the residuals as compared to the Quijote standard simulation output, which use different initial random seeds per simulation (See Figs. \ref{subfig:planck_spine_standard_residuals_appendix}, \ref{subfig:planck_standard_hmcode_residuals} and \ref{subfig:planck_standard_halofit_residuals}). For the fixed simulations, a similar residual pattern is consistently observed at scales of $k\sim0.1\;h\;\mathrm{Mpc}^{-1}$ across all models examined, whereas it is absent in the standard simulation residuals. This supports the scatter at intermediate scales as being consistent with sample variance in the common fixed-phase LH simulations, rather than an issue with any particular emulator.

\label{lastpage}
\end{document}